\documentclass[a4paper,fleqn,usenatbib]{mnras}

\usepackage{newtxtext,newtxmath}
\usepackage[T1]{fontenc}       
\usepackage{ae,aecompl}

\usepackage{graphicx}	% Including figure files
\usepackage{amsmath}	% Advanced maths commands
\usepackage{amssymb}	% Extra maths symbols

\title[A jet model for the IR variability of GX 339-4]{A jet model for the fast IR variability of the black hole X-ray binary GX~339-4}

\author[J. Malzac et al.]{Julien Malzac,$^{1}$\thanks{E-mail: julien.malzac@irap.omp.eu}
Maithili Kalamkar,$^{2}$
Federico Vincentelli,$^{2,3,4}$
Alexis Vue,$^{1}$
\newauthor
Samia Drappeau,$^{1}$
 Renaud Belmont,$^{1}$
Piergiorgio Casella,$^{2}$
Ma\"{i}ca Clavel,$^{5}$
\newauthor
Stéphane Corbel,$^{6}$
Micka\"{e}l Coriat,$^{1}$
 Damien Dornic,$^{7}$
Jonathan Ferreira,$^{5}$
Gilles Henri,$^{5}$
\newauthor
Thomas J. Maccarone,$^{8}$
Alexandre Marcowith,$^ {9}$
Kieran O'Brien,$^{10}$
Mathias P\'eault,$^{1}$
\newauthor
Pierre-Olivier Petrucci,$^{5}$
J\'erome Rodriguez,$^{6}$
David M. Russell,$^{11}$
 Phil Uttley$^{12}$\\
$^{1}$IRAP, Universit\'{e} de Toulouse, CNRS, UPS, CNES, Toulouse, France.\\
$^{2}$ INAF-Osservatorio Astronomico di Roma, Via Frascati 33, I-00040 Monteporzio Catone, Italy.\\ 
$^{3}$ DiSAT, Universitá degli Studi dell'Insubria, Via Valleggio 11,I-22100 Como, Italy. \\
$^{4}$ INAF - Osservatorio Astronomico di Brera Merate, via E. Bianchi 46, I-23807 Merate, Italy. \\
$^{5}$ Univ. Grenoble Alpes, CNRS, IPAG, F-38000 Grenoble, France.\\
$^{6}$ Laboratoire AIM, UMR 7158, CEA/CNRS/Universit\'e Paris Diderot, CEA DRF/IRFU/DAp, 91191 Gif-sur-Yvette, France\\
$^{7}$ CPPM, CNRS/IN2P3 - Universit\'e de M\'editerran\'ee, 163 avenue de Luminy, 13288 Marseille Cedex 09, France\\
$^{8}$ Department of Physics and Astronomy, Texas Tech University, Lubbock, TX, 79409-1051, USA \\
$^{9}$ Laboratoire Univers et particules de Montpellier, Universit\'e Montpellier/CNRS, place E. Bataillon, cc072, 34095 Montpellier, France\\
$^{10}$ Department of Physics, Durham University, South Road, Durham, DH1 3LE, UK\\
$^{11}$ New York University Abu Dhabi, PO Box 129188, Abu Dhabi, UAE\\
$^{12}$ Astronomical Institute Anton Pannekoek, University of Amsterdam, Science Park 904, 1098XH Amsterdam, Netherlands.\\
}

\date{Accepted XXX. Received YYY; in original form ZZZ}

\pubyear{2018}

\begin{document}
\label{firstpage}
\pagerange{\pageref{firstpage}--\pageref{lastpage}}
\maketitle

% Abstractof the paper
\begin{abstract}

Using the simultaneous Infra-Red (IR)  and X-ray light curves obtained by Kalamkar et al. (2016), we perform a Fourier analysis of the IR/X-ray timing correlations of the black hole X-ray binary (BHB) GX 339-4. The resulting IR vs X-ray Fourier coherence and lag spectra are similar to those obtained in previous studies of GX 339-4 using optical light curves. In particular, above 1 Hz, the lag spectrum features an approximately constant IR lag of about 100 ms. 
We model simultaneously the radio to IR Spectral Energy Distribution (SED), the IR Power Spectral Density (PSD), and the coherence and lag spectra using the jet internal shock model ISHEM assuming that the fluctuations of the jet Lorentz factor are driven by the accretion flow. It turns out that most of the spectral and timing features, including the 100 ms lag, are remarkably well reproduced by this model.  The 100 ms time-scale is then associated with the travel time from the accretion flow to the IR emitting zone. Our exploration of the parameter space favours a jet  which is at most mildly relativistic ($\bar{\Gamma}< 3$),  and a linear and positive relation between the jet Lorentz factor and X-ray light curve  i.e. $\Gamma(t)-1\propto L_{X}(t)$. 
 The presence of a strong Low Frequency Quasi Periodic Oscillation (LFQPO) in the IR light curve could be caused by jet precession driven by Lense-Thirring precession of the jet-emitting accretion flow. Our simulations confirm that this mechanism can produce an IR LFQPO similar to that observed in GX 339-4.

\end{abstract}


%%%%%%%%%%%%%%%%%%%%%%%%%%%%%%%%%%%%%%%%%%%%%%%%%%

%%%%%%%%%%%%%%%%% BODY OF PAPER %%%%%%%%%%%%%%%%%%

\section{Introduction}
\begin{figure}
	\includegraphics[width=\columnwidth]{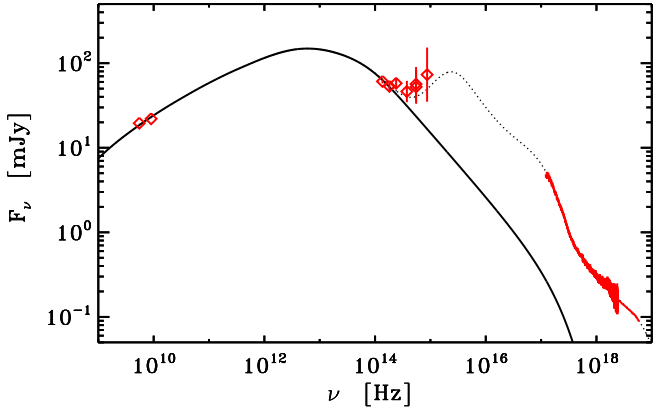}
   \caption{Comparison of the jet SED predicted by model A  (see Table~\ref{tab:par}) to the radio IR, optical  and X-ray measurements of K16 (in red). The synthetic SED represents an average of the simulated jet emission over 100 ks, it is shown by the solid curve. The model shown by the dotted curve adds the contribution from a self-irradiated accretion flow that was obtained from a fit of the data with the {\sc diskir} model (see Appendix~\ref{sec:diskir} and Table~\ref{tab:pardir} ). Both data and model are de-absorbed. { The de-reddened IR and optical fluxes are obtained using standard interstellar extinction law (Cardelli et al. 1989) with $A_V$ of 3.25 (Gandhi et al. 2011; K16). In the X-rays we used our best fit hydrogen column density $N_{H}=6.1\times 10^{21}$ cm  (see Appendix~\ref{sec:diskir})}. }
 \label{fig:fidsed}
\end{figure}

GX 339-4 is a recurrent black-hole X-ray binary transient which is known to exhibit fast sub-second variability (broad band noise and Quasi periodic Oscillations, QPO) over a broad multi-wavelength range from X-rays to Optical and Infra-Red (OIR) bands (Motch,  Ilovaisky \& Chevalier 1982; Gandhi et al. 2010, hereafter G10; Casella et al. 2010 hereafter C10; Motta et al. 2011; Gandhi et al. 2011; Kalamkar et al. 2016 hereafter K16, Vincentelli et al. 2018).
The aperiodic  X-ray variability is now generally believed to be caused by inwardly propagating fluctuations in the accretion flow  (Lyubarskii 1997; Arevalo \& Uttley 2006), while the X-ray QPOs could be related to relativistic precession (Stella \& Vietri 1998; Ingram 2016 and reference therein).
 In general, the fast OIR variability of black hole X-ray binaries can be caused either by reprocessing of the variable X-ray emission by the outer disc (O'Brien et al. 2002),  or by the variable synchrotron emission of the X-ray corona (Fabian et al. 1982; Merloni, Di Matteo \& Fabian 2000), or possibly a combination of both (Veledina, Poutanen \& Vurm 2011; Poutanen \& Veledina 2014; Veledina et al. 2017).  
 Alternatively, it could arise from variable synchrotron emission in the jet (Kanbach et al. 2001; Hynes et al. 2003;  Malzac et al. 2003; Malzac, Merloni \& Fabian 2004). The variable OIR emission  would then originate from the base of the jet emitting region at an elevation of $10^{3}$--$10^{4}$ gravitational radii\footnote{The gravitational radius of a black hole of mass $M$ is defined as  $R_{\rm g}$=$GM/c^2$, where $G$ is the gravitational constant and $c$ the velocity of light.} above the black hole (Malzac 2013; Malzac 2014, hereafter M14; Gandhi et al.  2017).
 
 In GX 339-4, the jet interpretation of the OIR variability is strongly favoured (G10; C10;  K16).  
Indeed, the OIR light curves are weakly correlated with the X-ray band and the OIR band lags behind the X-rays by about 100 ms (G10; C10). The OIR variability and its time response to the X-ray are too fast to be caused by disc reprocessing. The OIR lags are also quite difficult to explain in terms of synchrotron emission in the corona. In the synchrotron emitting hot flow model of Veledina et al. (2011), one would expect the opposite behaviour  (X-rays lagging behind OIR), or no lag at all. This model also predicts an anti-correlation between the X-ray and optical light curves. { Although such an anti-correllation is observed in some sources (e.g. XTE J118+480 or Swift J1753-0127), in the case of GX 339-4,} the observations show a positive correlation (at high Fourier frequencies at least). The 100 ms time lag is equally difficult to reconcile with a scenario in which both the OIR and X-rays are synchrotron emission produced by the jet (e.g. Markoff, Falcke \& Fender 2001), because, in this case, the optically thin synchrotron  emission in IR would be produced by the same population of leptons as the X-rays, although in this case the lag could be related to the cooling time-scale of the relativistic electrons (see discussion in C10).

On the other hand, the 100 ms lag can be naturally associated with the travel time between the accretion flow (producing the X-rays) and the OIR emitting region in the jet (Kanbach et al. 2001; Hynes et al. 2003; Eickenberry et al. 2008; G10, C10, K16, Gandhi et al.  2017). Hence, the OIR variability features of GX 339-4 have been attributed to jets. If this interpretation is correct, the study of the correlated X-ray and OIR timing in this source may allow us to probe the dynamics of accretion ejection coupling and test jet models.  

So far, the only detailed time-dependent emission model for compact jets in X-ray binaries is the internal shock model (Jamil, Fender \& Kaiser 2010; Malzac 2013; M14). As in similar models, developed in the context of gamma-ray bursts (Rees \& Meszaros 1994; Daigne \& Mochkovitch 1998; Bo{\v s}njak, Daigne, \& Dubus 2009) and AGNs (Rees 1978; Spada et al. 2001), it postulates fluctuations of the velocity of the jet. These variations of the jet velocity are generated by the central engine and then propagate along the jets. At some point, the fastest fluctuations start catching up and merging with slower ones. This leads to shocks in which a fraction of the bulk kinetic energy of the shells is converted into internal energy. Part of the dissipated energy goes into particle acceleration, leading to synchrotron and Compton emission.  The shape of the jet  Spectral Energy Distribution (SED) is almost entirely determined by the choice of the Fourier Power Spectral Density (PSD) of the jet velocity fluctuations (Malzac 2013; M14). 

The jet variability is most likely driven by the variability of the accretion flow which, in turn, is best traced by the X-ray light curves. The fluctuations of the jet Lorentz factor are thus expected to have a PSD that is similar to that of the X-ray light-curves.  In Drappeau et al. (2015, hereafter D15), we showed that an observed radio-IR SED of GX339-4 is matched by the model provided that the PSD of the jet Lorentz factor fluctuations has the same shape and amplitude as the simultaneously observed X-ray PSD. Similar results were obtained by P\'eault et al. (2018) who showed that the model could reproduce the evolution of the jet SED during an outburst of the black hole X-ray binary  MAXI~J1836-194.  Drappeau et al. (2017)  also suggested that the quenching of the radio emission observed in the soft state of BHBs could be related to the observed drop in X-ray variability (compared to hard state) which strongly reduces the jet radiative efficiency (see however Koljonen et al. 2018). The model is also supported by the observational results of Din\c{c}er et al. (2014) which indicate that black hole X-ray binaries with weaker X-ray variability in the hard state tend to be quieter in radio. 

The internal shock mechanism is intrinsically time dependent. Besides spectral shapes, the model naturally predicts a strong wavelength dependent variability  that resembles the observed one (M14; D15). Interestingly the PSD of the Lorentz factor fluctuations determines both the variability and spectral properties of the source. The spectral and timing properties are therefore intrinsically connected.  However the timing properties of the model remain to be studied in detail.
This is the main purpose of this paper in which we will  compare the predictions of the internal shock model to the fast IR and X-ray timing observations of GX 339-4 from K16.  The IR band is better suited than the optical to study and model the jet response to X-ray fluctuations,  as it is less likely to be contaminated by variability from the accretion flow (both from reprocessing and synchrotron). Indeed, any component arising from the accretion flow is stronger at shorter wavelength while the jet synchrotron emission is weaker.

The data analysis is presented in section~\ref{sec-obs-data-analysis}. In addition to the IR and X-ray power spectra which were already presented in K16 we calculate the Fourier coherence and phase/time lags spectra of an X-ray and IR light curve. In section~\ref{sec:fidmo}, our fiducial model is presented and its predictions are compared to the observed Spectral Energy Distribution (SED), PSD, coherence and lag spectra. In section~\ref{sec:precession}, the effects of jet precession are included in the model, and the consequences regarding the formation of IR QPOs are discussed. Finally, in section \ref{sec:parspace}, we investigate the effects of the model parameters on the timing predictions.

\begin{figure}
	\includegraphics[width=\columnwidth]{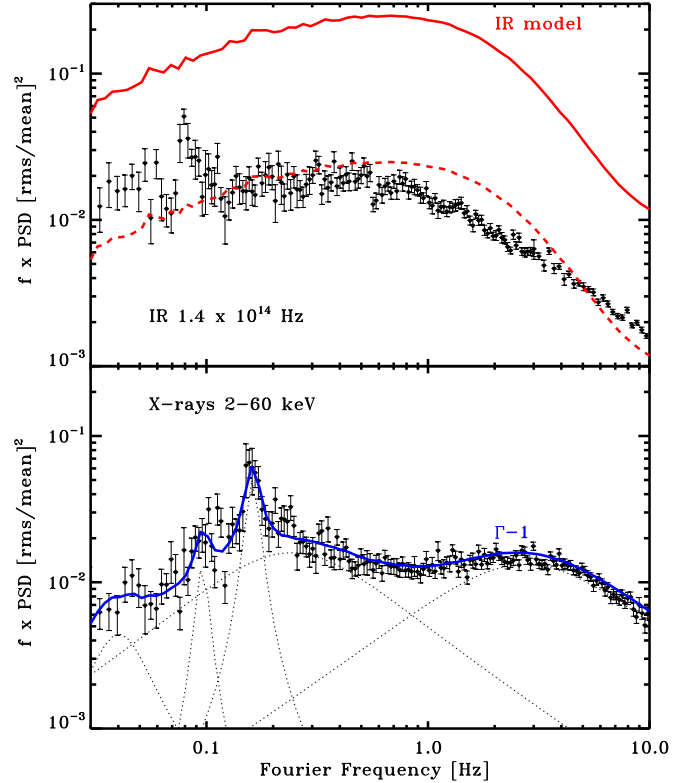}
   \caption{ Comparison of the PSD of model A (see Table~\ref{tab:par}) to the observation of K16. The black symbols shows the measured X-ray (bottom panel) and IR PSD (higher panel). Bottom panel: The dashed lines show the Lorentzian components of the best fit model to the X-ray PSD (as in K16).  The solid curve shows  the PSD of the jet Lorentz factor fluctuations which were generated using the best fit model of the X-ray PSD. Top panel: The solid curve shows the resulting model IR PSD, while the dashed curve shows the same model but with normalisation reduced by a factor of 10 to match the data. \label{fig:fidpsd}}
\end{figure}

\begin{figure}
	\includegraphics[width=\columnwidth]{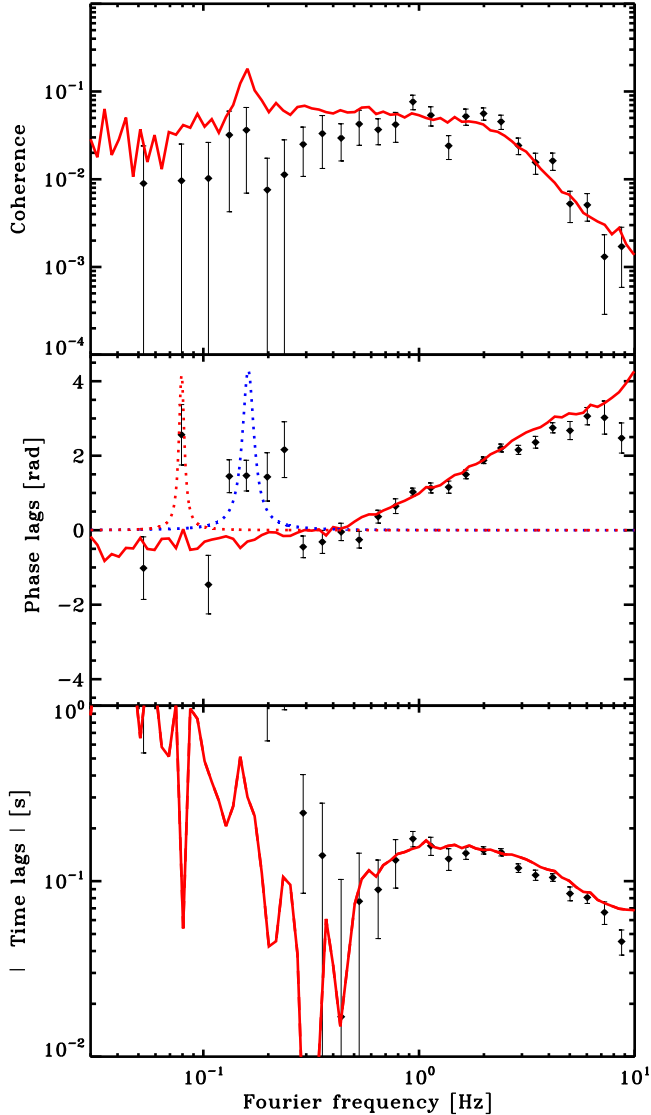}
   \caption{
   Comparison of the X-ray vs IR coherence and lags spectra of model A (see Table~\ref{tab:par}) to the observation of K16. The red curve shows model coherence (top), phase lag (middle panel) and time lags (bottom). The data are shown in black.  { The dotted curves in the middle panel show the best-fit Lorentzian profile for the main  IR (red) and X-ray (blue) QPOs plotted with arbitrary absolute, but exact relative, normalization.}
    \label{fig:fidccs}}
\end{figure}

 \begin{figure*}
	\includegraphics[width=\textwidth]{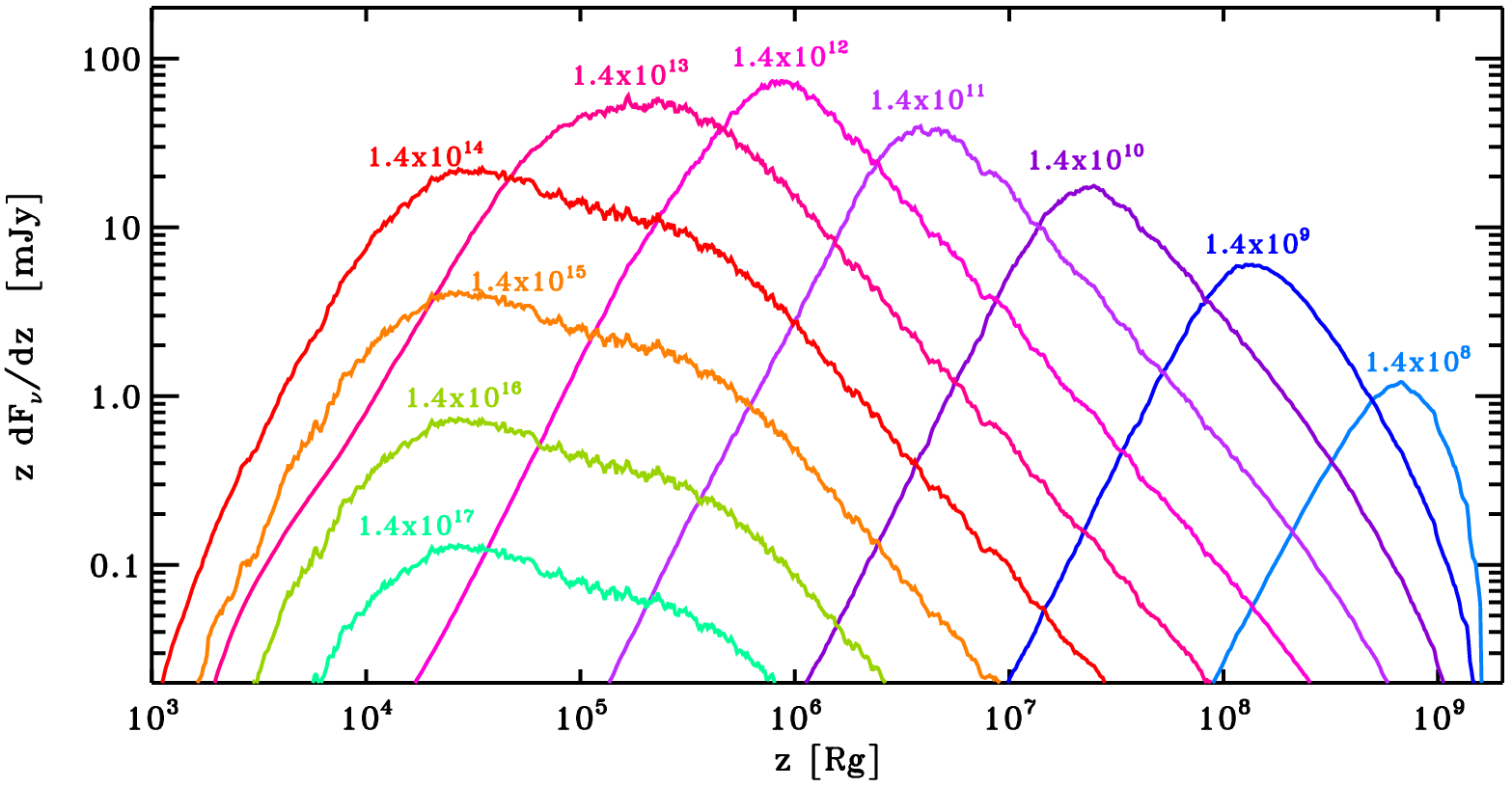}
   \caption{ Jet emissivity profiles across the electromagnetic spectrum. Time averaged emissivity profiles from model A at various photon frequencies ranging from $1.4\times10^{8}$ to $1.4\times10^{17}$ Hz as labelled. The band used in K16 corresponds to $1.4\times 10^{14}$ Hz.}
     \label{fig:lambdaprof}
\end{figure*}

\section{Observations and data analysis}\label{sec-obs-data-analysis}
 In this work, we use the observations of GX 339--4 obtained on March 28 2010 (MJD 55283) in the hard state during the rise of its outburst in 2010-2011 \citep[see e.g.,][]{cadolle2011-339, nandi2012-339, corbel-2013-339, dincer2012-339}. The source was observed in the IR K band using Very Large Telescope/Infrared Spectrometer And Array Camera (ISAAC), in two optical filters U and V with Optical Monitor on board \textit{XMM-Newton}, in X-rays with \textit{XMM-Newton} EPIC-pn and with the Rossi X-ray Timing Explorer (RXTE) Proportional Counter Array. {  These data were obtained simultaneously and at high time resolution allowing the study of fast variability. We refer the reader to K16 for the details of the data analysis. Quasi-simultaneous observations were reported in the radio band with ATCA \citep{corbel-2013-339}, and H,J,I,V bands with SMARTS \citep{buxton-339-2012}. The observed radio to X-ray SED is displayed on Fig.~\ref{fig:fidsed}.

{ The unabsorbed 0.1--200 keV flux is about 3~$\times$~10$^{-8}$~erg/cm$^{2}$/s. At the distance of 8~kpc and for a black hole mass of 10 solar masses which we will adopt in this paper (see  section~\ref{sec:ishem}) this corresponds to an isotropic luminosity of 2.3 $\times$10$^{38}$ erg/s, or equivalently, 18~percent of the Eddington luminosity, $L_{\rm E}$=1.3~$\times$~10$^{38}$$M_{\rm BH}/M_{\odot}$~erg/s. }

The detection of the first QPO in the IR band in a black-hole X-ray binary  was reported in this data set by K16. QPOs in the two optical bands (V and U) were also reported  at the same frequency as the infra-red QPO ($\sim$0.08 Hz). Interestingly, these QPOs were at half the frequency of the observed X-ray QPO \citep[classified as type-C;][]{wijnands1999qpos, casella2005} at $\sim$0.16 Hz; a weak sub-harmonic close to the IR and optical QPO frequency was also reported in the X-rays. The power spectra also showed the presence of broad-band noise components. It can be decomposed into three broad Lorentzian components (and two QPOs) in the X-ray band and two broad components (and one QPO) in the IR band (see K16). The X-ray and IR PSDs are displayed in Fig.~\ref{fig:fidpsd}. 

Using the same light curves as reported in K16 for the power spectral studies, we perform cross spectral analyses to compute the coherence and phase lags between the IR band and X-rays (from the RXTE data); due to poor statistics the optical/X-ray coherence and phase lags are poorly constrained and hence not presented here. 
{ The coherence function provides a measure of the degree of linear correlation between two time series as a function of Fourier frequency, while the argument of the complex cross spectrum provides the phase lag between the two time series (Bendat \& Piersol 1986, Nowak et al. 1999, Uttley et al. 2014).  All cross spectral products were computed using the procedures described in Uttley et al. (2014). In particular, following Vaughan \& Nowak (1997), we estimated the intrinsic coherence to take into account the Poissonian noise contribution. The computation was made using 1024 bin per segment (total length T=~38 s) and a logarithmic rebinning factor of 1.2. }

The coherence is shown in the top panel of Fig.~\ref{fig:fidccs}. It is weak and flat but significantly above zero between 0.3-2 Hz and falls steeply, { together with  the power spectrum}, above 2 Hz. The coherence below 0.3 Hz is weak and has large uncertainties.  The unconstrained coherence below 0.3 Hz occurs in conjunction with two different Lorentzian components dominating below 0.3 Hz in the power spectrum, which may be related to the loss of coherence (Vaughan \& Nowak 1997). Similar shape and strength of the coherence function was reported in this source by G10 in a decaying state (in 2007) in the optical/X-ray bands. An optical QPO was also reported in this observation, but without a simultaneous detection in X-rays. In both cases, the coherence value at the QPO frequency is consistent with the broad-band noise continuum. 

The phase lags between the IR/X-ray bands are shown in the middle panel of Fig.~\ref{fig:fidccs}. A positive value indicates an IR lag against X-rays. As the lags are defined between -$\pi$ to +$\pi$, the lags jump to -$\pi$ once these become higher than +$\pi$. The corresponding time lags can be calculated by dividing the phase lag by 2$\pi f$, which are shown in the bottom panel of Fig.~\ref{fig:fidccs}. The IR phase lag increases smoothly in the range 0.3-6 Hz, which can be associated with the observed slowly decreasing time-lag which keeps an amplitude of the order of 100 ms in this frequency range. 
 {  The phase lags at 0.05 Hz and 0.1 Hz, lie on the 0.3-6 Hz lag extrapolation (constrained by large errors) and  appear to switch to a negative value. Similar behaviour at low frequencies was reported in this source in the optical/X-ray bands by G10. However, at and around the optical/IR and X-ray QPO frequencies, the lags appears to be positive  which is observed for the first time. This indicates a connection between the QPOs where the X-ray QPO (at 0.08 Hz) leads the IR QPO. Moreover, although no IR QPO is detected at 0.16 Hz, a positive phase lag is also observed at that frequency. } \\

\begin{figure*}
	\includegraphics[width=\textwidth]{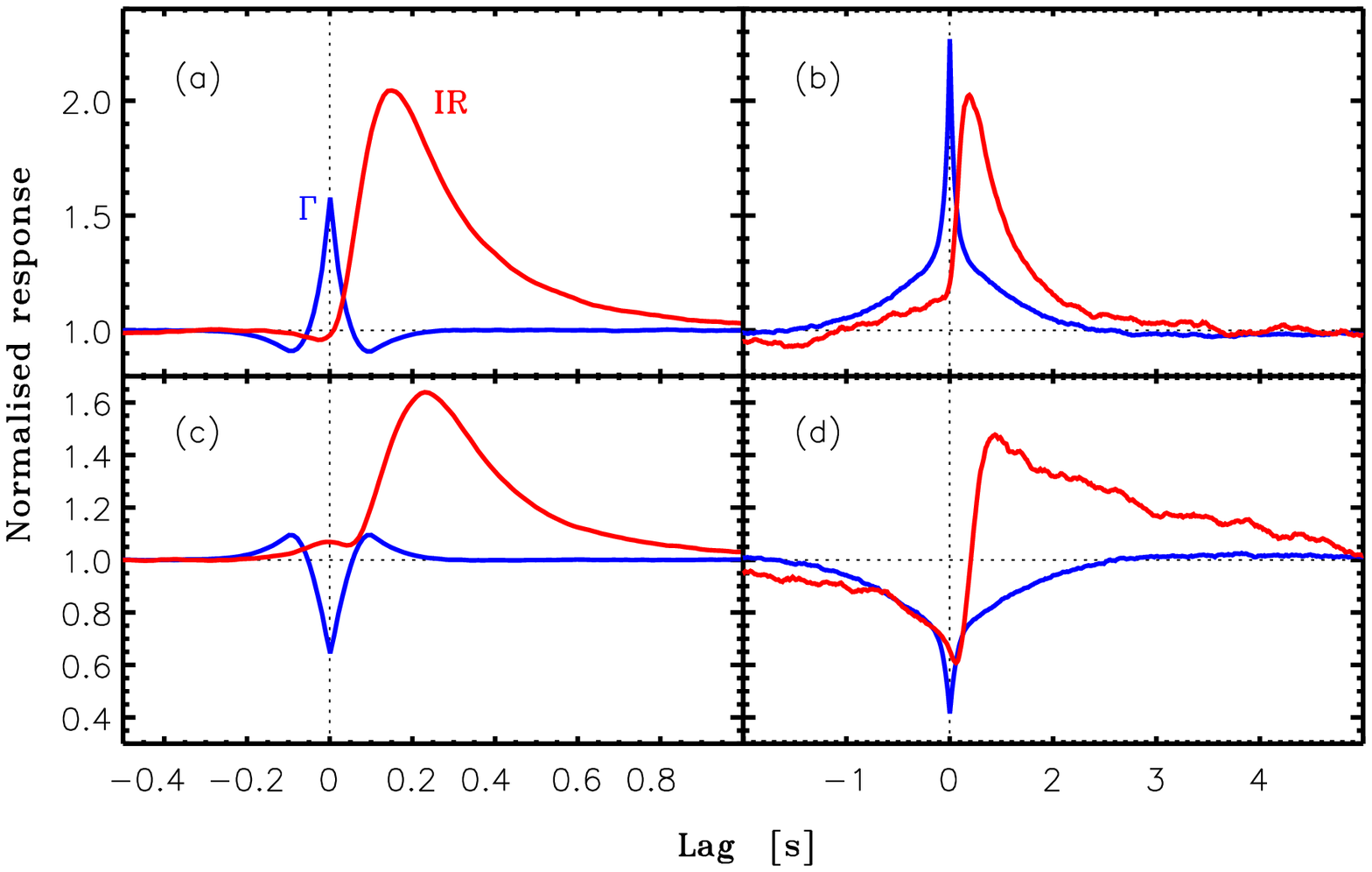}
   \caption{ Examples of averaged synthetic IR time responses to spikes (panels a and b)  and dips (panel c and d) of the jet Lorentz factor $\Gamma$. Panels (a) and (c) show the IR responses to short events, while panel (b) and (d) show the response to longer events. In each panel, the blue curve shows the average dip or spike selected in the the time series of  $\Gamma-1$  from model B (see Table~\ref{tab:par}), while the red curve stands for the corresponding  average IR response.    These results were obtained using the event superposition technique described in section~\ref{sec:flares}.  We set the  selection threshold   $f_s=1.4$ in all panels. The selection time-scale is  $t_s=0.2$ s for panel (a) and (c), and  $t_s=20$ s for panel (b) and (d).\label{fig:flares}
}
\end{figure*}

\section{Modelling}\label{sec:fidmo}

\subsection{\sc ishem}\label{sec:ishem}
We use  {\sc ishem}, the numerical implementation of the internal shock model described in M14. In this model, 
{ the continuous jet is approximated as a collection of a large number of uniform small-scale ejecta}.
Homogenous shells of gas are continuously ejected at the base of the jet with variable Lorentz factors and at { uniform} time-intervals comparable  to the dynamical time-scale of the inner accretion flow ({ $9.941$ ms for the simulations shown in this paper}). { Unless specified otherwise, we assume that the shells are all ejected with the same mass.} The ejecta travel along the jet and expand according to the fixed half-opening angle $\phi$ of the  conical jet. Due to their different velocities they may collide and merge with other ejecta.  The code follows the propagation and  hierarchical merging of the shells.  Supersonic shell collisions lead to the formation of shocks. During the duration of the shock crossing time,  a fraction of the kinetic energy is gradually converted into relativistic leptons and magnetic field according to fixed equipartition factors.  The lepton energy distribution is assumed to be a power-law energy distribution with index $p$, $n_e(\gamma) \propto  \gamma^{-p}$ ,  for lepton  Lorentz factor $\gamma$ in the range,  $\gamma_{\rm min}< \gamma <\gamma_{\rm max}$, with $\gamma_{\rm min}$, $\gamma_{\rm max}$ and $p$ being fixed parameters.  The time dependent synchrotron emission of the shocked ejecta is calculated taking into account the expansion of the shells. The energy losses due to adiabatic expansion are taken into account but radiation losses are neglected. Inverse Compton emission may lead to gamma-ray emission, but does not contribute in the radio to IR frequency range. It is neglected in the current version of {\sc ishem}.

We follow the approach of Drappeau et al. (2015, 2017) which assumes that the variability of the jet is driven by the accretion flow which in turn can be traced by X-ray light curves. In practice, we generate fluctuations of the jet Lorentz factor which have the same PSD as the observed X-rays, both in shape and amplitude of variability.  For the present study, the fluctuations of the jet Lorentz factor are generated according to the best fit multi-Lorentzian model to the RXTE power spectrum obtained in K16 (and shown Fig.~\ref{fig:fidpsd}). 
This entirely determines the shape of the model SED. The other parameters of the models such as the time averaged jet kinetic power  $\bar{P}_{\rm J}$, the jet half-opening angle $\phi$, average jet Lorentz factor $\bar{\Gamma}$, and inclination angle between the jet axis and the line of sight  $i$, only allow the SED to be shifted in frequency or in normalisation, but do not affect its shape. 
Throughout this paper and unless specified otherwise, for the purpose of comparison we use the same parameters as in D15. In particular, the mass of the black hole is set at 10~$M_{\odot}$, the inclination angle $i=23^{\circ}$ at a distance of 8~kpc. These values were adopted by D15 because they were both in agreement with the existing published observational constraints on mass and inclination, and also allowed to fit the SED of GX339-4 with reasonable {\sc ishem} parameters (see discussion and references in D15). { We note however that a recent near-IR study by Heida et al. (2017) suggests a lower mass (2.3 $M_{\odot}$< $M_{\rm BH}$< 9.5 $M_{\odot}$) and a larger inclination in GX 339-4    (37$^{\rm o}$<i<78$^{\rm o}$) . A lower mass has no impact on the result of our modelling, it changes only the estimate of the Eddington Luminosity and therefore the jet power and X-ray luminosity would be higher when expressed as Eddington fractions. The effects of a larger inclination will be illustrated in section~\ref{sec:incl}.  We note also that the jet is not necessarily perpendicular to the orbital plane of the binary system and therefore the jet inclination that we use in {\sc ishem} may differ from the orbital inclination constrained by spectroscopy of the donor star.} 

Following D15 we also assume that during shell collisions, half of the dissipated energy is converted into relativistic leptons and the remaining into turbulent magnetic field. 

{ We list the models considered in this work in Table~\ref{tab:par}. Model A is our basic model discussed in section~\ref{sec:sed}. In  Model A' the parameters are modified so that the IR emission is dominated by the outer disk instead of the jet  (see section~\ref{sec:psd} and Appendix~\ref{sec:diskir}). Model B and the following other models, assume that the jet precesses at the frequency of the IR QPO and that the X-ray QPO does not contribute to the variability of the jet Lorentz factor (see section~\ref{sec:precishem}). Models C1 to C4 are similar to model B but consider variations of the jet inclination angle $i$ (see section~\ref{sec:incl}). Models D1 to D5 consider variations of the jet Lorentz factor (section~\ref{sec:ejlf}). Models E1 to E3 explore different prescriptions for the relation between the X-ray light curve and jet Lorentz factor (see section~\ref{sec:jdc}),  while model F1 to F4 are similar to the E models but assume that the ejected shells have a variable mass and a constant kinetic energy. In all models except A',  the jet half-opening angle $\phi$ and kinetic power are  adjusted so that the SED reproduces the observed radio and IR fluxes.}

 \begin{table}
 \caption{Model parameters: }
 \label{tab:par}
 \begin{tabular}{cccccccc}
  \hline \\
  Model & $\bar{P}_{\rm{J}}/L_{\rm E}$ & $\phi$ (deg) & ${\bar{\Gamma}}$ & $i$ (deg) & Q   &  g &   Figures   \\ 
  \hline \\
 A &9.5$\times 10^{-2}$    &    2.3     &     2          &     23        &  0   &  1 & \ref{fig:fidsed}--\ref{fig:lambdaprof} \\
 A' & 0.30                        &     20       &     2          &    23       &    0    &  1    &  \ref{fig:fidsedbigshift}  \\
 B   &0.15                        &    3.5     &     2          &     23        &  10 &    1 &    \ref{fig:fidcspec}--\ref{fig:gm0}     \\
 C1   &1.19                        &     3.4        &   2      &      85       &  10   &  1 & \ref{fig:incl}     \\
 C2   &0.61                        &     3.5        &    2     &      60        &   10  &  1 & \ref{fig:incl}   \\
 C3   &0.29                        &     3.5          &   2    &      40         &   10  &  1 &\ref{fig:incl}  \\
 C4   & 8.2  $\times 10^{-2}$    &      3.5          &  2       &      1          & 10     &  1 &  \ref{fig:incl} \\ 
 D1    & 0.53                         &     55      &     1.1     &     23        &  10  &   1 &  \ref{fig:gam},\ref{fig:gamprof}          \\
 D2   &  0.23                       &       11        &    1.5     &    23        &   10  &   1 &    \ref{fig:gam},\ref{fig:gamprof}          \\
 D3   &   0.12                        &       0.99    &     3      &    23         &   10  &   1 &  \ref{fig:gam},\ref{fig:gamprof}  \\
 D4   &   0.18                         &     0.17      &      6     &    23         &    10  &  1   &   \ref{fig:gam},\ref{fig:gamprof}  \\
 D5   &  0.37                       &        0.05     &    10    &     23        &   10  &    1 &  \ref{fig:gam},\ref{fig:gamprof}       \\
E1    & 0.15                 &  3.5              &     2      &      23         & 10     & -1 & \ref{fig:gm0} \\
E2    & 0.25                    & 1.1                & 2            &   23              & 10    & 0.5 & \ref{fig:gm0} \\
E3    &  8.4$ \times10^{-2} $ &  10.7          &   2          &   23               & 10   &   2 &  \ref{fig:gm0} \\
F1    &     0.15      &  3.8 & 2 & 23 & 10 & 1 &   \ref{fig:gm0} \\
F2  &     0.14      &  3.8 & 2 & 23 & 10 & -1 &   \ref{fig:gm0} \\
F3  &     0.25     &  1.1  & 2 & 23 & 10 & 0.5 &   \ref{fig:gm0} \\
F4  &  8.1 $\times10^{-2}$  &  13.4          &   2          &   23               & 10   &   2 &  \ref{fig:gm0} \\

  \hline
 \end{tabular}
\end{table}

\subsection{Spectral energy distribution}\label{sec:sed}

{ When comparing the spectral predictions of {\sc ishem} with an observed SED, there are two possibilities:
\begin{itemize}
\item[(i)] The observed X-ray PSD used for the input fluctuations in {\sc ishem} leads to an SED shape that is incompatible with the observations: then, there is no tuning of the other {\sc ishem} parameters that will allow to reproduce the observation. The model cannot account for the data (at least not using the X-ray PDS as input). 
\item[(ii)] The resulting predicted SED has a shape that can match the data: then, due the to strong degeneracy of the model parameters, there are many different combinations of the {\sc ishem} parameters which allows one to fit the SED equivalently. Therefore the best fit model parameters cannot be uniquely determined from SED fitting.  However most of the model parameters have to fulfil additional constraints originating from independent observations (e.g. measurement of distance or orbital plane inclination)  or physical considerations  (e.g. jet opening angle cannot be too large; jet power is unlikely to be super-Eddington in low luminosity sources).  Then demonstrating that there is at least one 'reasonable' combination of {\sc ishem} parameters which allows to fit the SED and also complies with all the other observational and theoretical constraints, provides additional test of the model. This may also, in turn, tighten the constraints on the jet parameters. 
\end{itemize}

With the data set of K16 on GX 339-4, we are clearly in case (ii). The predicted synthetic SED (model A in Table~\ref{tab:par}) is compared to the radio and IR measurements  in  Fig.~\ref{fig:fidsed}.  Model A matches the data in radio and IR both in flux and spectral slopes. At higher frequencies there is an additional component from the accretion flow and the jet model alone cannot account for optical  UV and X-ray data.  Although this paper focusses on the synchrotron jet, for illustrative purposes, Fig.~\ref{fig:fidsed} presents a plausible accretion flow model ({\sc diskir} , Gierli{\'n}ski, Done, \& Page 2009) accounting for the high frequency part of the SED. Some details about the   model and overall  fit procedure are given in Appendix~\ref{sec:diskir} and the model parameters are shown in Table~\ref{tab:pardir}. The resulting reduced $\chi^2$ of the fit is slightly less than unity.  }

{ The parameters of model A were chosen as follows.}  D15 used data that were taken 17 days  (2010 March 11) before the observations considered in the present paper. They showed that the observed  SED was remarkably well reproduced by the model for reasonable values of the parameters.
Because we use a different PSD of the fluctuations than D15, and because  both radio and IR fluxes are  higher by a factor of about 2 with respect to the data used in D15, the preferred model parameters of D15 cannot be used to fit the current SED. In order to determine the best fit parameters we started with a simulation using the same model parameters favoured by D15.  We then calculate by how much this model SED must be shifted both in frequency and normalisation, in order to minimise the difference between the model and the radio and IR data points (see Appendix~\ref{sec:diskir}). The effect of each parameter on the model normalisation and typical frequencies is known analytically (Malzac 2013; M14). This can be used to determine a new set of parameters that shifts the model frequency and normalisation by the amount required to obtain the best possible match to the data. This procedure is detailed in P\'eault et al. (2018). One can then run a simulation in order to check that the new parameters give a good agreement to the data.

Since many different sets of parameters can give an equally good fit to the data, we first decided to allow the jet power and jet opening angle to differ from that of D15 and keep all the other parameters identical to that of D15. This resulted in very large jet power reaching about 0.5 $L_{\rm E}$  (to be compared to the X-ray luminosity $\simeq0.18 L_{\rm E}$) and a jet opening angle of about 10$^{\circ}$, much larger than the 1$^{\circ}$ assumed in D15. Such parameters are not excluded but appear somewhat extreme. We tried other combinations and finally decided to change the energy distribution of the synchrotron emitting electrons. The minimum Lorentz factor of the accelerated electrons was set  to $\gamma_{\rm min}=10$ (instead of $\gamma_{\rm min}=1$ in D15), and the electron power law distribution was set to $p=2.5$ (instead of $p=2.3$ in D15)  and we used the 'slow' shock dissipation scheme while D15 used the 'fast' one  (see M14  for a discussion of the 'slow' and 'fast' dissipation prescriptions). { $p=2.5$ corresponds to the best fit estimate from the IR optically thin slope obtained by Gandhi et al. (2011). Together, the higher $\gamma_{\rm min}$ and $p$ parameters allow us to fit the data with a reduced jet power (P\'eault et al. 2018). Indeed,} this leads to a best fit  jet opening angle of 2.35$^{\circ}$ and a jet power of 0.095 $L_{\rm E}$ which is  comparable to the X-ray  power and  therefore more in line with the typical value for compact jets at this level of X-ray luminosity.

We note that our approach assumes that the X-ray emission is dominated by the accretion flow. In order to self-consistently ensure that the jet synchrotron emission is negligible in the RXTE band, we introduced a posteriori  a cut-off to the optically thin jet synchrotron component. This cut-off , that we set at 1 keV, could be related to the highest energy of the accelerated electrons in the jet, or most likely mimic a radiation cooling break (which is not accounted for by the simple radiation transfer scheme of {\sc ishem} ). We stress that although the presence of such a  cut-off is debatable, it does not affect the predictions of the model in the radio to IR band  which are the prime focus of this paper (see discussions in Drappeau et al. 2017 and P\'eault et al. 2018).

{ Fig.~\ref{fig:lambdaprof} displays the jet emissivity profile for model A. At photon frequencies below 1.4 $\times 10^{13}$ Hz the jet radiates in the partially absorbed regime and the emission at a set photon frequency originates from a specific distance scale in the jet: longer wavelengths are produced at larger distance in the jet. Above  1.4 $\times 10^{13}$  Hz, the emission becomes mostly optically thin and originates from an extended region  comprised in the approximate range 10$^{4}$ --10$^{6}$ $R_{\rm G}$ with a maximum   emissivity around  around 2$ \times 10^{4}$ $R_{\rm G}$ . This corresponds to the range of distances probed by the IR light curves in this model.} 

\subsection{Power spectra} \label{sec:psd}
 { \sc ishem} also produces  synthetic  IR light curves with high time-resolution from which we can calculate the model IR PSD.  The resulting IR PSD for model A is displayed in Fig.~\ref{fig:fidpsd} and compared to the observed IR PSD. 
The overall shape of the synthetic IR PSD is roughly similar to the observed one. However the model predicts a much stronger IR variability amplitude than observed. As shown in Fig.~\ref{fig:fidpsd} the normalisation of model would have to be reduced by a factor of about 10 in order to match the data.  This implies that the rms/mean variability amplitude of the model is about 3 times that observed. As already suggested by M14 the model could be reconciled to the data by the presence of an additional constant flux component in IR that would damp the observed relative variability amplitude. In our case, this requires a constant component contributing to at least 70 percent of the observed flux in the K band. { A possible  origin of this component could be the accretion disc or the hot accretion flow. As noted above, in the optical at least, the observed SED seems to require a dominant disc component. In Appendix~\ref{sec:diskir} we show that the spectral data of K16 are statistically compatible  with disc dominated emission in the $K$ band (model A' of Appendix~\ref{sec:diskir} shown in Fig~\ref{fig:fidsedbigshift}) . However, it is very unlikely that such a disc dominated model could fit the SED compiled by Gandhi (2011) which has a much better IR coverage and was taken only 17 days before the SED of K16 and clearly shows a dominant non-thermal component dominating the K band.}  Another possibility could be that internal shocks are not the only dissipation mechanism leading to IR synchrotron emission in the jet. { For instance, in addition to internal shocks, one can not exclude the presence of a recollimation shock close to the base of the jet. Such a standing shock is likely to form in magnetically driven jets launched from an accretion disc as the hoop stress dominates centrifugal forces and forces recollimation toward the jet axis (Ferreira 1997). It could provide a steady energisation of the jet producing the additional constant component. This component would be significant only in IR. Indeed, the recollimation shock is expected to form close to the base of the emitting region where the particle density and magnetic field are such that the emission at longer wavelengths is self-absorbed.  The particles accelerated in this shock quickly cool down as they travel along the expanding jet, the emission produced at larger distances from the black hole (or equivalently at lower photon frequencies)  would remain dominated by the internal shock mechanism.} 
 Assuming the internal shock contribution represents only 27 percent of the IR flux and fitting for the jet power and jet opening angle keeping  all the other parameters at their fiducial value leads to  $\bar{P}_{J}=0.22 L_{\rm E}$ and $\phi=13.7^{\circ}$. 

\begin{figure}
	\includegraphics[width=\columnwidth]{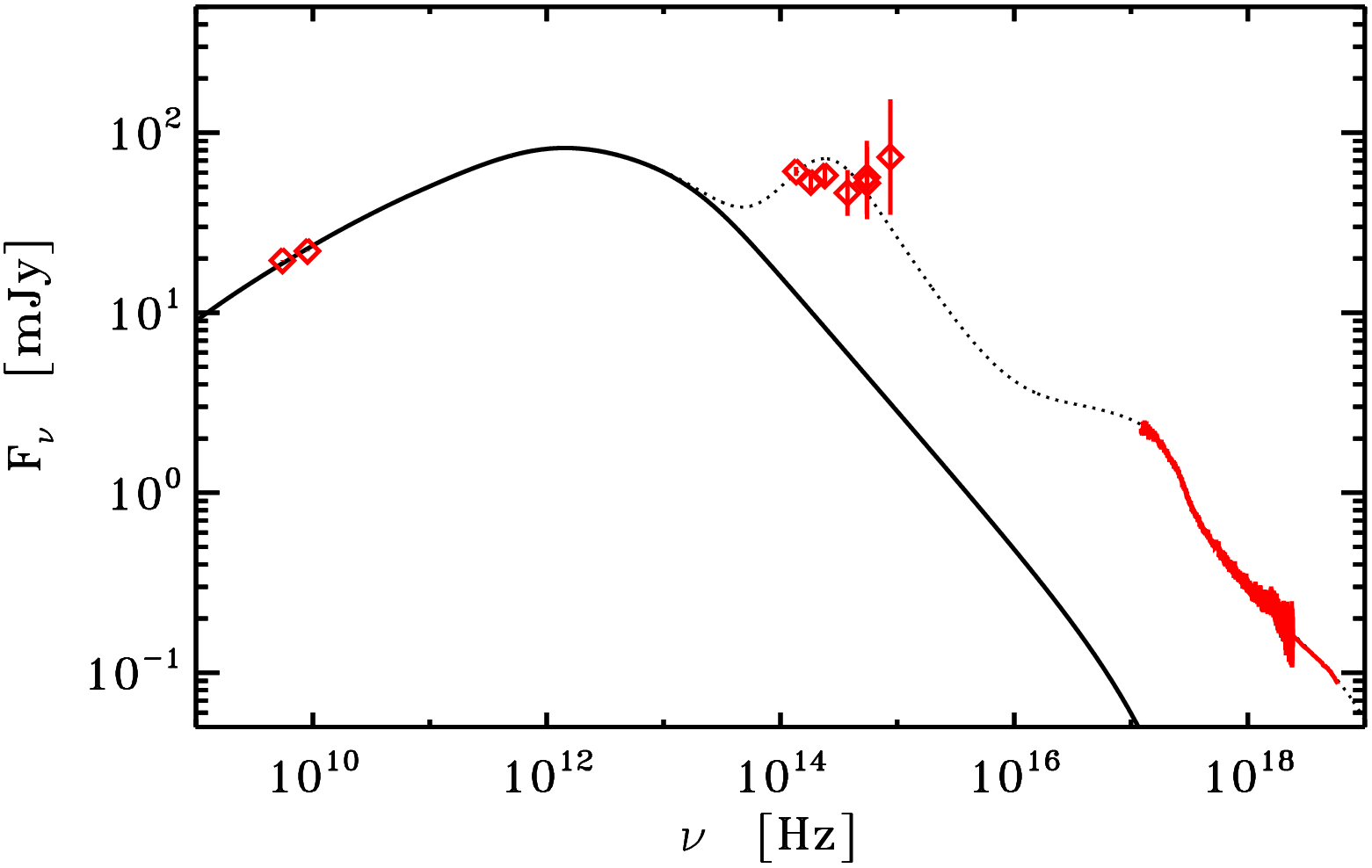}
   \caption{Comparison of the jet SED predicted by model A'  (see Table~\ref{tab:par}) to the radio IR, optical  and X-ray measurements of K16 (in red). The synthetic SED represents an average of the simulated jet emission over 100 ks, it is shown by the solid curve. The model shown by the dotted curve adds the contribution from a self-irradiated accretion flow that was obtained from a fit of the data with the {\sc diskir} model (see Appendix~\ref{sec:diskir} and Table~\ref{tab:pardir}). Both data and model are corrected for absorption. { The de-reddened IR and optical fluxes are obtained using standard interstellar extinction law (Cardelli et al. 1989) with $A_V$ of 3.25 (Gandhi et al. 2011; K16). In the X-rays we used the best fit hydrogen column density $N_{H}=5.2\times 10^{21}$ cm  (see Appendix~\ref{sec:diskir})} }
 \label{fig:fidsedbigshift}
\end{figure}

 We note also that radiative cooling is not implemented in the current version of {\sc ishem} and could affect strongly the predicted IR variability.  Indeed, although in general adiabatic expansion losses dominate  the cooling of the accelerated electrons in compact jets, the  IR emitting region overlaps with  the very base of the jet emitting region (located at a few thousand gravitational radii from the black hole see Fig.~\ref{fig:lambdaprof}) where the magnetic field is still strong enough so that synchrotron cooling may dominate over expansion losses (see e.g. Pe'er \& Casella 2009 and discussion in Chaty, Dubus \& Raichoor 2011). The implementation and study of the consequences of radiation cooling are out of the scope of this paper and are postponed to future works. One can speculate that radiation cooling may damp the amplitude of predicted IR variability down to a level closer to that observed.

{ Besides the PSD normalization,  the shape of the predicted  PSD is not completely  satisfactory.  As can be seen in Fig.~\ref{fig:fidpsd}, neither the break frequency, nor the slopes of the observed PSD are accurately reproduced. Better agreement with the data may be obtained by tuning the inclination and average jet Lorentz factor (see sections~\ref{sec:incl} and \ref{sec:ejlf}). Indeed these parameters control the amplitude of relativistic Doppler beaming effects that, in turn, affect the IR timing response (see section~\ref{sec:flares}). Moreover, the effects of radiation cooling, as well as the slowly variable additional component that appears required to dilute the predicted variability, may distort the shape of the IR PSD. The study of these effects and a detailed fitting of the IR PSD is postponed to future works.}

Another important issue is that the model PSD does not exhibit an IR QPO. The injected fluctuations of the jet Lorentz factor contain the X-ray  QPO and one could expect that feeding an oscillating $\Gamma$ to the jet would result in a similar oscillation of its IR emission.  However the injected QPO is not strong enough to produce a significant IR QPO.  A close inspection of the model PSD suggests a very weak feature in IR at the frequency of the X-ray QPO. This feature is much weaker than the observed IR QPO and even more importantly, it is not at the right frequency, since the observed X-ray and IR QPOs are in harmonic ratio.  Therefore the IR QPO is unlikely to result from  the dynamics of internal shocks. Instead the production of the IR QPO requires an additional ingredient to the model. For instance, the  IR QPO may arise from a contribution from the accretion flow (see Veledina, Poutanen \& Ingram 2011; Veledina \& Poutanen 2015). In the context of the jet model,  jet precession may lead to the formation of an IR QPO. This possibility will be investigated in details in section~\ref{sec:precession}.

 \begin{figure}
	\includegraphics[width=\columnwidth]{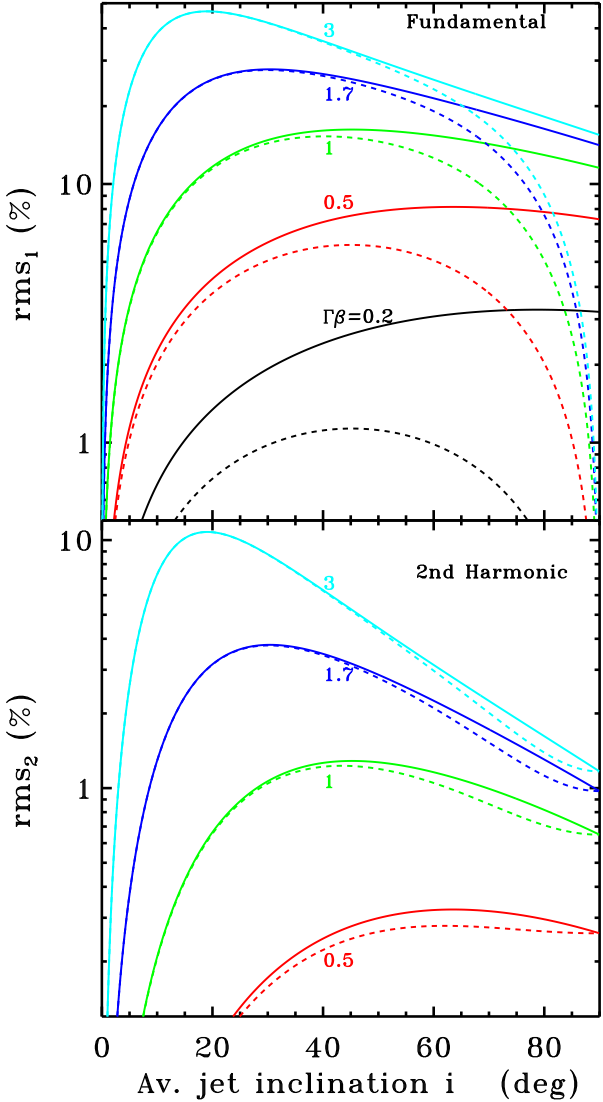}
   \caption{rms amplitude of the fundamental (top panel) and second harmonic (bottom panel) of the QPO versus inclination angle $i$ for $p=2.3$, $\psi=5^{\circ}$ for various jet reduced kinetic momenta $\Gamma\beta$ as labelled. The solid lines are the result for a one sided jet, the dashed lines assume that the counter-jet is visible }
     \label{fig:rmsvsi}
\end{figure}
 
\subsection{Synthetic IR response to fluctuations of the jet Lorentz factor}\label{sec:flares}

 { In our model  the radiative response of the jet to the fluctuations of $\Gamma$ is complex and strongly non-linear.  
 In order to understand this better, it is useful to consider the average IR response to specific events  in the time evolution of $\Gamma$. We can for instance, consider the response to a fast rise and then decrease of $\Gamma$ (a spike) or, on the contrary a fast decrease and the recovery of $\Gamma$ (a dip). 
To estimate this we  have applied an event superposition technique to our synthetic light curves.  
We use the time series  of the variations of $\Gamma$ to select spikes or dips that we stack together.  To select the events we define a threshold ratio $f_s$ and a time-scale of selection $t_s$. The spikes then are selected according to the following criteria:
The peak $\Gamma-1$ value of the spike is $f_s$ times the local value as obtained from an average over time $t_s$. The peak bin is further required to represent a maximum of $\Gamma$  over bins within $t_s/2$ before and after the peak bin. The selected spikes are then peak aligned and averaged. The corresponding pieces of synthetic  IR light curves are also centred on the peak time bin of $\Gamma-1$ and averaged in the same way. 
The dips of $\Gamma-1$ are selected and superposed and their IR response estimated in a similar way. The minimum value of the dip is $1/f_s$ times the local count rate as obtained from an average over time $t_s$ and represents a local minimum over bins within $t_s/2$ before and after. 

Fig.~\ref{fig:flares} displays some examples of averaged dips and flare profiles and their respective IR response.  
 It shows  that the average response to a spike of  $\Gamma$ is an IR flare caused by the faster than average shells launched during the spike sweeping the jet. In this case the IR flare is delayed by a time related to the travel time of the fast shells before they start catching up with the other ones, the duration of the IR flare may be much longer than the duration of the $\Gamma$ spike. Interestingly,  the response to a fast dip in $\Gamma$  will lead to a similar IR flare.  In other words the response to a dip is negative.  This is because the slower shells that have been injected during the dip will soon be the targets of the faster shells that are injected after the dip.  In this case however the IR response is broader and its peak is significantly more delayed  than in the case of the response to a spike. Moreover, if the $\Gamma$ dip is long enough the IR flare can be preceded by an IR dip because the continuously decreasing velocities in the first phase of the dip temporarily switches-off shell collisions. Overall, because the IR flux level depends on the difference of Lorentz factors of the colliding shell,  on long times scales IR flux response corresponds roughly to the time derivative of the jet Lorentz factor.}
  
{ However, the variations of $\Gamma$  also induce a modulation of the light curve through Doppler beaming and this may affect the IR response. This effect is usually weak compared to the intrinsic shock variability. However, the Doppler effects depend on the inclination and the time-averaged jet Lorentz factor, $\bar{\Gamma}$. Different choices of these parameters can change dramatically  the amplitude and time-scale of the IR response.}

Finally, we note that the strong QPOs that are present in the temporal evolution of $\Gamma$ for model A interferes with the spike and dip selection method and induces spurious effects on the resulting IR response. For this reason the results presented in Fig.~\ref{fig:flares} were obtained from a simulation in which the X-ray QPOs are removed from the input fluctuations of $\Gamma$ (model B see section~\ref{sec:precishem} and Table~\ref{tab:par}).   
 
\subsection{X-ray vs IR coherence and lags}\label{sec:coherenceandlags}

So far we have assumed that the input jet Lorentz factor fluctuations have the same PSD as the X-ray emission. This does not necessarily require that the jet fluctuations are correlated with the X-rays. Since the observations indicate significant correlations and lags between the X-ray and IR bands and since, in the model, the IR variability is driven by the fluctuations of the jet Lorentz factor, there must be a relation between the X-ray light curve and  Lorentz factor fluctuations. This relation certainly  depends on the physics of the dynamical coupling between accretion and ejection, which is essentially unknown.  
In Fig.~\ref{fig:fidccs}, the synthetic X-ray vs IR coherence and lags have been calculated assuming that the X-ray luminosity scales linearly with the jet Lorentz factor, i.e.:
\begin{equation}
L_{\rm X} \propto \Gamma-1. \label{eq:xvsg}
\end{equation}
A more general model  including fluctuations of the shell masses and a non-linear connection of the  X-ray light curve to the fluctuations of $\Gamma$  will be explored in section~\ref{sec:parspace}. 

We note that the phase lags are determined modulo $2\pi$ and since the lags are presumably small, the range between $-\pi$ and $+\pi$ is used to define the observed phase and time lags. However  occasionally and at some Fourier frequencies the model can predict lags that are comparable or longer than the time-scale of the fluctuations.  When this happens the phase lag flips from   $+\pi$  to $-\pi$, leading to discontinuities and sometimes  strong oscillations in the phase-lag spectra. This also results in a time-lag that does not correspond to the physical time-scale predicted by the model. 
For the purpose of clarity of the figures, in this paper, the models lag spectra are calculated assuming a continuous lag spectrum and allowing for phase lag values outside the range [$-\pi$,+$\pi$]. The synthetic lags are first determined within [-$\pi$, +$\pi$] in the same way as the observed lags, then the potential discontinuities are removed by adding or subtracting  $2\pi$  to the phase lags in the frequency range of interest. This explains for example why the model phase lags are greater than $\pi$  at the highest frequencies in Fig.~\ref{fig:fidccs}.

Overall we find that the predictions of the model are in remarkable agreement with the observations at high Fourier frequencies. 
The drop in coherence at high frequencies as well as the shape of the lag spectra is well reproduced.  The nearly constant 100 ms time lag, above $\sim$0.5 Hz is related to the travel time between the accretion flow and  the IR emitting region.  Taking into account the projection effects, an observed travel time of $\tau=100$ ms  corresponds to distance along the jet of $z_{\rm lag}\simeq  \tau \bar{\beta} c / (1- \bar{\beta}\cos{i})\simeq 9\times 10^3 R_{\rm g}$, where $\bar{\beta}=\sqrt{1-\bar{\Gamma}^{-2}}$. We note that the peak of  the time average IR emissivity profile in the jet model is slightly  farther away, around  $2\times10^4 R_{\rm g}$ (see Fig.~\ref{fig:lambdaprof}). Indeed  the lag depends not only on the location where the bulk  of the IR emission is produced but also where the bulk of the IR variability occurs i.e. the innermost region of the jet where the first collisions occurs is much more variable. This reduces the lag by a factor of a few compared to the naive expectations. 

At lower frequencies, the model predicts negative lags that are in qualitative agreement with the observations. 
{  These negative lags are caused mostly by  the long time-scale IR flares driven by dips of $\Gamma$ which dominate the correlation on long time-scales (because the IR response to dips is longer than the response to spikes, see Fig.~\ref{fig:flares}). The result is that the negative response to dips at positive lags translates into negative Fourier lags. Another way to see it is to consider that at zeroth order the IR flux on long time-scales corresponds to the time-derivative of the X-ray flux (see section~\ref{sec:flares}), so we expect a negative phase lag of -$\pi/2$ which is close to that obtained from the simulations  at long time-scales.}

The positive lag observed around the X-ray and IR QPO frequencies are most likely associated with the QPOs and are therefore not reproduced by this model.
We note however that the synthetic coherence is enhanced at the X-ray QPO frequency confirming that the oscillations of the Lorentz factor are partially transferred to the IR band although not sufficiently to produce a strong IR QPO feature.  
  
\begin{figure}
	\includegraphics[width=\columnwidth]{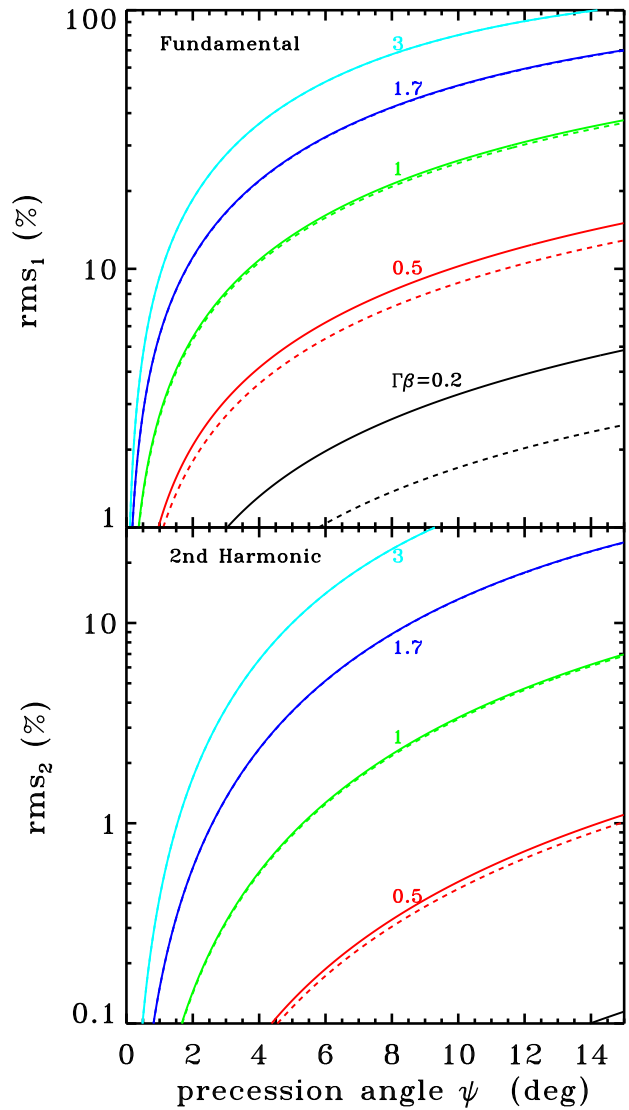}
   \caption{rms amplitude of the fundamental (top panel) and second harmonic (bottom panel) of the QPO versus precession angle $\psi$ for $p=2.3$, $i=25^{\circ}$ for various jet reduced kinetic momenta $\Gamma\beta$, as labelled. The solid lines are the result for a one sided jet, the dashed lines assume that the counter-jet is visible}
     \label{fig:rmsvspsy}
\end{figure}

\section{Jet precession model for the IR QPO}\label{sec:precession}

 Low Frequency Quasi-Periodic Oscillations (LFQPO) are often observed in X-rays at Fourier frequencies ranging from $10^{-2}$ to a few Hz and rms amplitude between 3 and 30 percent.  A popular model for the  X-ray  LFQPO  involves global Lense-Thirring (LT) precession of a hot accretion flow (Ingram, Done \& Fragile 2009).  LT precession is a frame dragging effect associated with the misalignment of the angular momentum of an orbiting particle and the BH spin, leading to precession of the orbit. 
Numerical simulations have shown that in the case of a hot geometrically thick accretion flow, this effect can lead to global precession of the hot flow (Fragile et al. 2007). The hot flow precesses like a solid body, and the precession frequency is given by a weighted average of the LT precession frequency between inner and outer radii of the flow. 
The emission of the precessing hot flow is then naturally modulated by a mixture of relativistic Doppler beaming, light bending and Compton anisotropy.  This model predicts the right range of observed LFQPO frequencies. The amplitude of the LFQPO depends on the details of the geometry and viewing angle. The rms is usually larger at high inclinations and can reach 10 percent (see Ingram et al. 2015 also for predictions of modulation of the polarisation of observed X-ray radiation). 

The IR QPO may be produced in the accretion flow. The precession of the hot flow may also lead to a modulation of its OIR synchrotron emission (Veledina, Poutanen \& Ingram 2013)  or modulation of illumination of the outer disc (Veledina \& Poutanen 2015)  possibly producing a QPO signal.  { However, because the correlation between the band limited IR and X-ray noises cannot be explained with these accretion flow models, jet precession appears to be a more likely explanation for the IR QPO of GX 339-4 (see discussion in K16).} Indeed, if the X-ray LF QPOs are caused by global LT precession of the hot flow and if the jet is launched from the accretion flow, one may expect the jet to precess with the flow. Recent GRMHD simulations of a tilted accretion flow and jet suggest that this is indeed the case (Liska et al. 2017). The (mostly) optically thin synchrotron radiation observed in IR and optical would then be modulated at the precession frequency due to the variations of Doppler beaming effects towards the observer.

Modelling the X-ray QPOs in the framework of the LT precession model requires the knowledge  of the angular distribution of the radiation emitted by the accretion flow. This depends on the size, geometry density and temperature profile of the Comptonizing region and other details which are essentially unknown. Previous studies have assumed a certain accretion flow geometry to estimate the QPO modulation profile, amplitudes and harmonic content (Veledina et al. 2013; Ingram et al. 2015). Our data set is very constraining for these models,  in particular because in our data, the X-ray QPO appears to be dominated by the second harmonic rather than the fundamental.  In the specific  model of Veledina et al. (2013), this is expected when the average angle between the line of sight and the direction perpendicular to the accretion flow is around 60$^{\circ}$.  However at these inclinations the predicted  QPO rms amplitudes are a few percents at most, significantly smaller than what is observed by K16 in GX 339-4. This appears to exclude the specific accretion flow structure used by these authors.  The IR QPO amplitude and lag with respect to the X-ray will provide further constraints. Developing a full model for the X-ray QPO that can be coupled with the IR jet model to constrain the geometry of the accretion flow is beyond the scope of this paper and is reserved for future work.
Instead, in the following, we will focus on the expected properties of the IR QPO  caused by jet precession.  In section~\ref{sec:anes} we will use simple analytical estimates to explore how  the amplitude of the IR QPO depends on the geometric parameters, in section~\ref{sec:counterjet} the effects of the counter-jet will be discussed, then in section~\ref{sec:precishem} we will present the results of {\sc ishem} simulations taking jet precession into account.  

\subsection{Analytical estimates}\label{sec:anes}

   From our modelling of the observed SED,  the IR jet synchrotron emission appears to be mostly in the optically thin regime (see Fig.~\ref{fig:fidsed}). For an optically thin  power-law spectrum emission of spectral slope $\alpha$ the modulation of the observed flux is given by  $F_{j} \propto \delta^{(2-\alpha)}$, where $\delta=[\Gamma(1-\beta\cos{i})]^{-1} $ is the usual relativistic Doppler boosting factor { (see e.g. equation A19 of M14)}.  In the case of optically thin synchrotron emission by relativistic electrons with a power-law energy distribution of index $p$, i.e. $n_{e}(\gamma)\propto \gamma^{-p}$, $\alpha=(1-p)/2$. We assume that the jet precesses at a frequency $f_{\rm p}$ around an axis which makes an angle $i$ with the line of sight. The  amplitude of the precession is given by the constant angle $\psi$ between the jet and the precession axis. The modulation is then given by: 
 \begin{equation}
  F_{\rm j} \propto \left(1-\bar{\beta} \mu\right)^{-\frac{3+p}{2}}, 
  \label{eq:fjmod}
 \end{equation}
 where $\bar{\beta} c$ is the average velocity of the jet (velocity fluctuations are neglected), and
 \begin{equation}
 \mu=\cos i\cos\psi+\sin i \sin\psi \cos\left(2\pi f_{\rm p}t\right)
 \label{eq:mudet}
 \end{equation}
  where $t$ is the lab frame time of emission and its origin is chosen so that the jet direction, the precession axis and the line of sight all lie in the same plane at $t=0$. 
  { In equation~\ref{eq:fjmod}, the factor  $1/\Gamma$, appearing in the expression of $\delta$, was omitted as it does not affect the dependence of the flux on the viewing angle.}  
  
   The resulting QPO rms amplitude is very sensitive to the parameters and a wide range of QPO amplitudes can be produced. 
 Fig.~\ref{fig:rmsvsi} shows the rms amplitude of the QPO as a function of inclination angle for different jet velocities. For a face on jet ($i=0$) there is no modulation as the jet is always seen with the same angle. The amplitude of the QPO increases with the inclination (at least at low $i$). As expected the amplitude of the QPO increases sharply with the jet velocity.  If the jet is relativistic the rms reaches a maximum (which is about 40 percent for $\Gamma \beta=3$) before decreasing slowly with inclination.  The second harmonic has an amplitude that is about a factor of 10 lower than the fundamental. As shown in Fig.~\ref{fig:rmsvspsy}, the QPO is also very sensitive to the precession angle $\psi$ and its fractional rms amplitude increases very quickly with $\psi$ and can quickly reach values that are larger than 100 \% in the fast  jet case.  
We note that the amplitude of the QPO could be strongly reduced, if a non-jet constant component is present in IR. The level of constant flux inferred from the modelling of the IR PSD of GX 339-4 (see Sect.~\ref{sec:fidmo}) would reduce the fractional rms by a factor of $\sim$3.

\subsection{Effect of counter jet}\label{sec:counterjet}

  Depending on the geometry of the binary system and accretion flow, the counter jet may be visible, adding a contribution to the modulation. We assume that the jet and counter jet are symmetric and have the same temporal emission pattern. We note that the counter jet emission lags behind that of the jet because of the longer photon travel time to the observer. However, the jet IR emitting regions are close to the black hole. The observed X/ IR lags indicate that it takes only $\sim$100 ms for the information to travel from the inner part of the accretion flow to the IR emitting region. Photon travel time delays should be at most of this order and can be safely neglected against the $\sim$ 10 s  time-scale of the precession. The modulation profile can then be approximated as: 
  \begin{equation}
  F_{j} +F_{\rm cj}\propto \left(1-\bar{\beta} \mu\right)^{-\frac{3+p}{2}}+\left(1+\bar{\beta} \mu\right)^{-\frac{3+p}{2}}
 \end{equation}

\begin{figure}
	\includegraphics[width=\columnwidth]{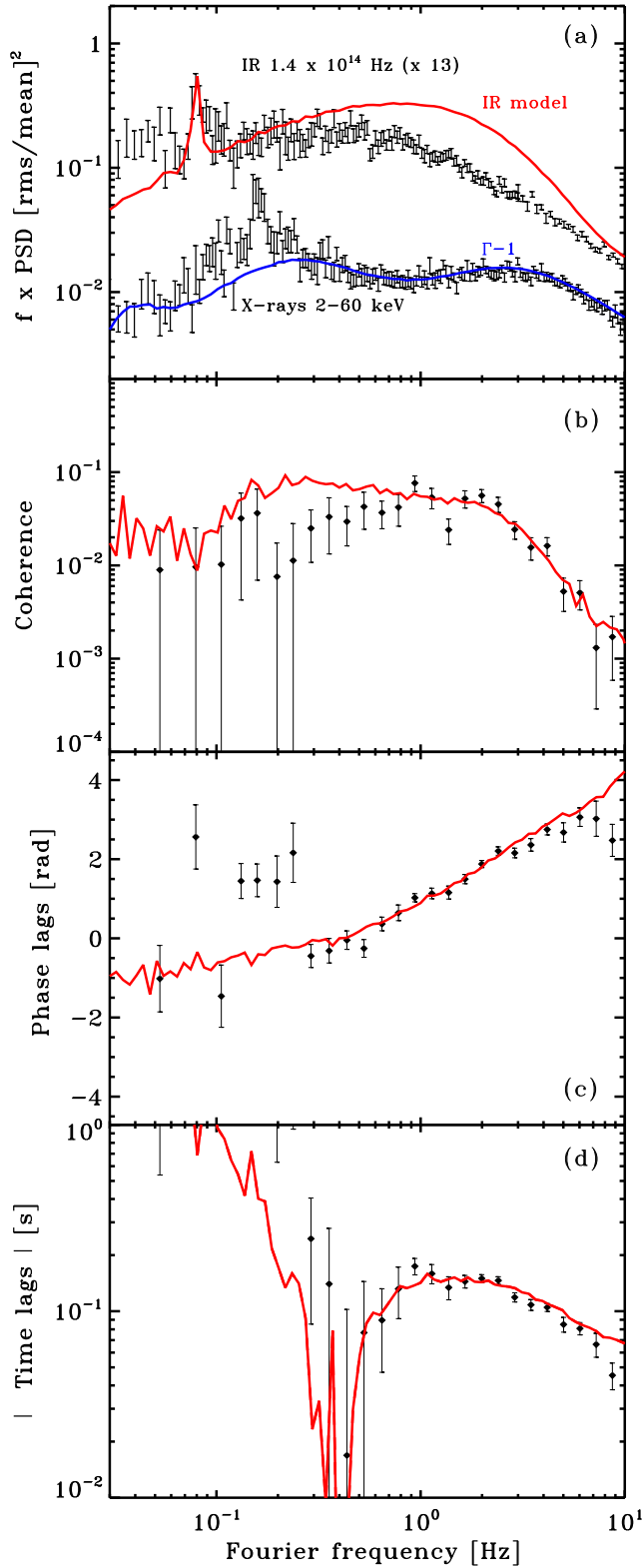}
   \caption{Panel (a):  Comparison of the PSD of model B to the observation of K16. The blue curve shows the X-ray PSD which also corresponds to the PSD of the jet Lorentz factor fluctuations with QPO features subtracted. The red curve shows the model IR PSD. The data are shown in black. The normalisation of the observed IR PSD is multiplied by a factor of 10. The other panels display a comparison of the X-ray vs IR coherence and lags spectra of the fiducial model to the observation of K16. The red curve shows the model coherence (panel b), phase lags (panel c) and absolute value of time lags (panel d). The data are shown in black. }
     \label{fig:fidcspec}
\end{figure}

\begin{figure}
	\includegraphics[width=\columnwidth]{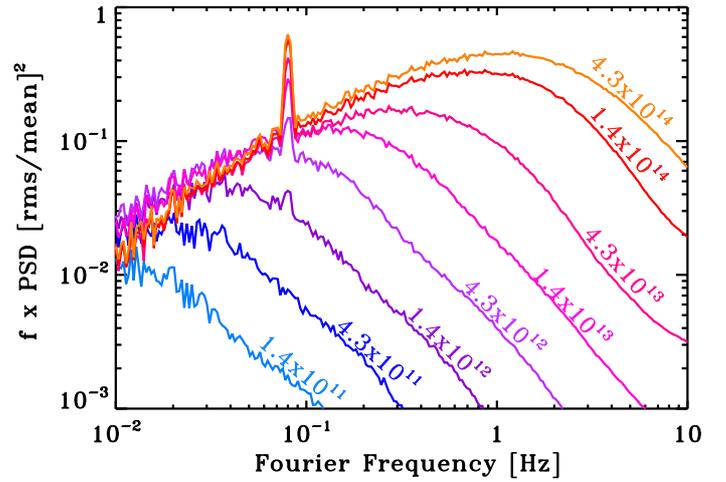}
   \caption{ Dependence of the synthetic PSD on the photon frequency. The parameters are those of the fiducial model B also shown in Fig.~\ref{fig:fidcspec}.  The various curve shows the synthetic PSD calculated at photon frequencies of $1.4\times10^{11}$ Hz, $4.3\times10^{11}$ Hz, $1.4\times10^{12}$ Hz, $4.3\times10^{12}$ Hz, $1.4\times10^{13}$ Hz, $4.3\times10^{13}$, $1.4\times10^{14}$ Hz, and  $1.4\times10^{14}$ Hz as labelled.}
     \label{fig:psdnu}
\end{figure}

The resulting QPO amplitude of the two-sided jet model is also shown on Figs.~\ref{fig:rmsvsi}  and~\ref{fig:rmsvspsy}.  { At low inclination and large Lorentz factor, the contribution from the counter-jet is negligible due to Doppler beaming effects. At large inclination and lower jet velocity, the counter-jet component becomes comparable to that of the jet but shifted in phase by half a precession period.}
  As the jet and counter jets emission are anti-phased, the effect of the counter-jet is then to reduce the amplitude of the oscillation which goes down to zero in the fundamental for edge-on inclinations.  Interestingly the effect is much weaker in the second harmonic which can become dominant at large inclinations. { This is because a shift by half a precession period corresponds to a full period of the second harmonic}.
  Overall, the effects of the counter jet is significant only in slow jets or at  large inclination, when doppler boosting is minimal.

  Moreover, the IR emitting region of the counter-jet is likely to be obscured by the accretion flow. Its visibility depends on the elevation $z_{\rm IR}$ of the IR emitting region, on the size of the accretion disc $R_{\rm d}$ and the viewing angle $i$. The counter-jet is visible only if  $i<tan^{-1}(R_{\rm d}/z_{\rm IR})$.
  In GX 339-4, the orbital measurements of Heida et al. (2017) indicate $R_{\rm d}\sim3\times 10^{11}$~cm.
   For an elevation of $z_{\rm IR} \sim 10^4 \, R_{\rm G} \sim 1.5 \times 10^{10}$~cm, visibility of the counter-jet  would  require $i>87^{\circ}$, which is larger than all current estimates of the orbital inclination of GX 339-4. Alternatively, if the disc is truncated in its inner parts, the IR emitting region of the jet could be visible through this central hole in the accretion disc. Assuming that the IR radiation is not absorbed or scattered in the hot inner flow and ignoring light bending effects, visibility requires a viewing angle $i<tan^{-1}(R_{\rm in}/z_{\rm IR})$.  During the observations considered in this paper, the source was in a bright hard state in which the inner radius of the disc, $R_{\rm in}$ cannot be very large (see e.g. Plant et al. 2015; De Marco et al. 2017). Assuming $R_{\rm in}< 100 \quad R_{\rm G}\simeq1.5\times  10^{8}$~cm would require $i<0.6^{\circ}$ to be able to see the counter jet. This is smaller than all current estimates of the orbital plane inclination.  It is therefore very unlikely that the counter jet of GX 339-4 is visible at IR wavelength and from now on we will consider only the emission from the jet pointing toward the observer. 

\subsection{Jet precession in {\sc ishem}}\label{sec:precishem}
 
In the context of the internal shock model, the randomly variable velocity of the jet as well as the dynamics of shell collisions could significantly reduce the amplitude of the oscillations.  In order to investigate this issue, jet precession was implemented in {\sc ishem}.  The main features of the numerical model are the following:

 We assume each shell is ejected in a slightly different direction, according to the precession direction at the time of ejection, and then propagates ballistically. When two shells collide their 3D momenta are added up so that the resulting shell then travels in a direction that is closer to the precession axis. In this process the precession is gradually damped along the jet as the hierarchical merging of the ejecta takes place.  
  At large distances, where the jet is constituted mostly by the product of collisions of many shells that were ejected over a time $\ga 1/f_{\rm p}$, the radial component of the velocity has vanished and their trajectory is almost exactly along the precession axis. { Therefore, precession occurs only close to the base of the jet and does not have a significant effect on the jet opening angle measured at large distance e.g. from radio measurement. It does not lead to a jet opening angle much larger than observed as could be expected if the whole jet precessed with the hot flow.}    
  
{  In this exploratory version of the model, the trajectory of the ejecta is followed only in 1 D along the z axis. In this scheme, we consider that a collision occurs exactly as in the non-precessing model when two shells reach the same elevation. We do not consider the amount of 3D overlap of the shells or indeed the possibility that an incoming ejecta may 'miss' its target due to too different trajectories.  This however is a reasonable approximation as long as the precession angle is not too large compared to the jet opening angle.} 

 In the simulations presented in this paper, we assume that the IR emitting region of the counter jet is hidden from our view, only the emission from the jet pointing towards the observer is taken into account. 
 
 The modulation predicted by equation~\ref{eq:mudet} is perfectly periodic and would imply that the QPO profile in the PSD is a delta function. However, the observed X-rays and IR QPOs both have a significant width. This implies that the amplitude, the frequency and/or the phase of the oscillations vary in time. In the context of  the LT precession model this could in principle be caused by modulations of the precession angle or frequency  driven by fluctuations of the mass accretion rate and accretion flow spin axis.  However to our knowledge, those coherence breaking mechanisms have never been investigated in detail and such a study is clearly out of the scope of this paper. In order to account for these effects in a simple manner, we introduce a new parameter $Q$, which defines the number of cycles over which the precession remains coherent. In the simulations, the phase of the precession is changed to a new random value after every $Q$ cycles.
 
 Also, since in our scenario both the X-ray and IR QPOs are caused by the geometrical effects associated with precession of the hot flow and the jet, the X-ray oscillation should not be fed to the jet. For this reason and from now on, the X-ray QPO is subtracted from the power spectrum used to generate the fluctuations of $\Gamma$. In practice we use the same multi-Lorentzian best fit model of the X-ray PSD as before but with the two QPO features  at ~0.09 and 0.16 Hz removed (see Fig.~\ref{fig:fidpsd}).  This new PSD  of the fluctuations of $\Gamma-1$ is shown in the top panel of Fig.~\ref{fig:fidcspec} and compared to the X-ray PSD.  As the shape of the SED is sensitive to the input PSD of the fluctuations, we had to change slightly the jet power and jet opening angle  in order to ensure that the radio and IR fluxes are matched by the model. The new model parameters are  $\phi=3.5^{\circ}$  and $P_j=0.145 L_E$ respectively. In the following we refer to this model as model B. 

 Fig.~\ref{fig:fidcspec} also shows the IR PSD predicted by {\sc ishem} for a precession frequency $f_{\rm p}=0.08$~Hz  and angle $\psi=5^{\circ}$ and $Q=10$.  
 One can see that the model IR QPO has an amplitude that is qualitatively consistent with the observations (once the correction for the constant component is applied). This QPO amplitude is also in agreement with the simple analytical estimates presented above.    
Fig.~\ref{fig:psdnu} illustrates the dependence of the synthetic PSD on wavelength.  As already discussed in M14 the jet behaves as a low-pass filter gradually removing the fastest variability at longer wavelength. { Moreover, as can be seen on Fig.~\ref{fig:psdnu}, and as expected, the amplitude of the QPO quickly decreases at longer wavelengths due the  damping of the radial component of the shell velocities.    At 4.3$\times 10^{11}$ Hz there is no trace of a QPO in the PSD. From Fig.~\ref{fig:lambdaprof} we see that this wavelength is emitted mostly around an elevation of $\sim 10^{6} R_{\rm G}$. This implies that  the precession is already completely damped at this scale and is therefore unlikely to have a significant effect on the radio jet structures observed on the much larger scales of $10^7$--$10^9 R_{\rm G}$.}

The effect of varying $Q$ is illustrated in Fig.~\ref{fig:Q}, as $Q$ is increased the QPO becomes stronger and narrower because it is more coherent, the data suggest $Q\sim10$. {  From a fit of the observed IR PSD we find that the quality factor of the observed IR QPO  is $Q$=14$\pm3$, which is comparable. }

  Fig.~\ref{fig:fidcspec} also displays the  X-ray vs IR coherence and lag spectra. They are similar to those of the fiducial model A (see Fig.~\ref{fig:fidccs}) despite the slightly different PSD of fluctuations used as input.  
The absence of low frequency oscillation in the input fluctuations of  $\Gamma-1$ reduces the coherence at low frequency, making it closer to the observed coherence.  The model lags are slightly more negative at low frequencies.  The positive lags measured in the data at the IR and X-ray QPO frequencies are most likely related to the coupling between X-ray and IR QPO and are not expected to be reproduced by the model since the X-ray QPOs are  not taken into account.

\begin{figure}
	\includegraphics[width=\columnwidth]{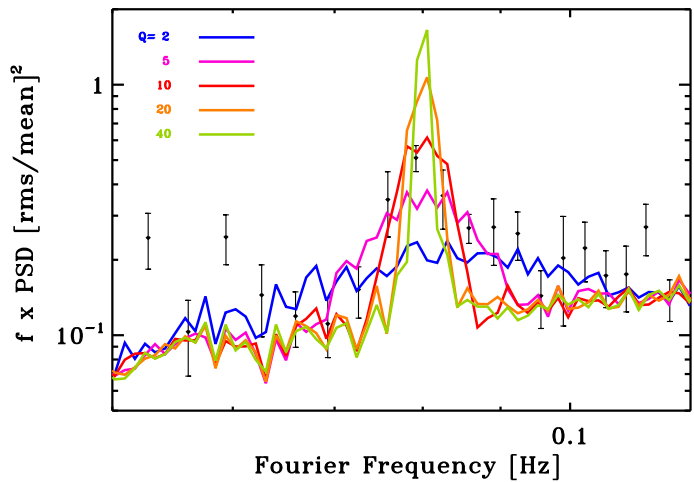}
   \caption{ Dependence of the synthetic QPOs on the quality factor parameter $Q$. The other model parameters are those of the fiducial model B also shown in Fig.~\ref{fig:fidcspec}.  }   \label{fig:Q}
\end{figure}

\section{Exploring parameter space}\label{sec:parspace}

As mentioned above the main driver of the spectral shape is the power spectrum used as input for the jet Lorentz factor fluctuations. The other parameters like the  average jet Lorentz factor, the jet opening angle, inclination or the  jet kinetic power, change only the normalisation of the SED and shift the SED shape along the frequency axis. This shift of the SED along  the frequency axis  induces changes in the timing properties observed at a fixed frequency. For instance let us suppose that  the jet kinetic power is increased while all the other parameters (including $\bar{\Gamma}$) are kept constant. The model predicts that the SED shifts towards higher frequency because the energy density in the jet increases. However changing the jet kinetic power has negligible effects on the shell collisions and the (normalised) shock dissipation profile along the jet is practically unchanged.  The only difference is that our reference band (let us say the IR band) now probes larger distances in the jet. Therefore  we will observe longer time-scales in the IR PSD and lags. In the end, in terms of timing properties, the effect of changing the jet power and observing at a fixed frequency is almost equivalent to observing the jet at different frequencies (as in Fig~\ref{fig:psdnu}). The same can be said for most of the model parameters such as the jet opening angle for example.  Indeed, when these parameters are varied the fixed observed frequency will probe radiation coming from different regions of the jet and this will affect the timing properties.

 Let us now consider a different set of these parameters that produce exactly the same model SED. This is the situation that we have with GX 339-4: we have a fixed observed SED that the model can fit with many different combinations of parameters. 
 In general,  the changes in the different parameters combine so that the IR band always probes the same region of the jet and the IR timing properties are unchanged. In other words, there is  degeneracy not only in the spectral but also in the timing  properties.
 
This is not true of all the parameters however. As we will show in the following, the inclination angle and the average jet Lorentz factor have a deeper effect on the timing properties that can be used to break some of the degeneracies.  Therefore,  by using simultaneously the timing and spectral data one may constrain not only the basic jet parameters but also the dynamical accretion-ejection coupling processes.  

\subsection{Effects of the inclination angle}\label{sec:incl}
\begin{figure}
	
	\includegraphics[width=\columnwidth]{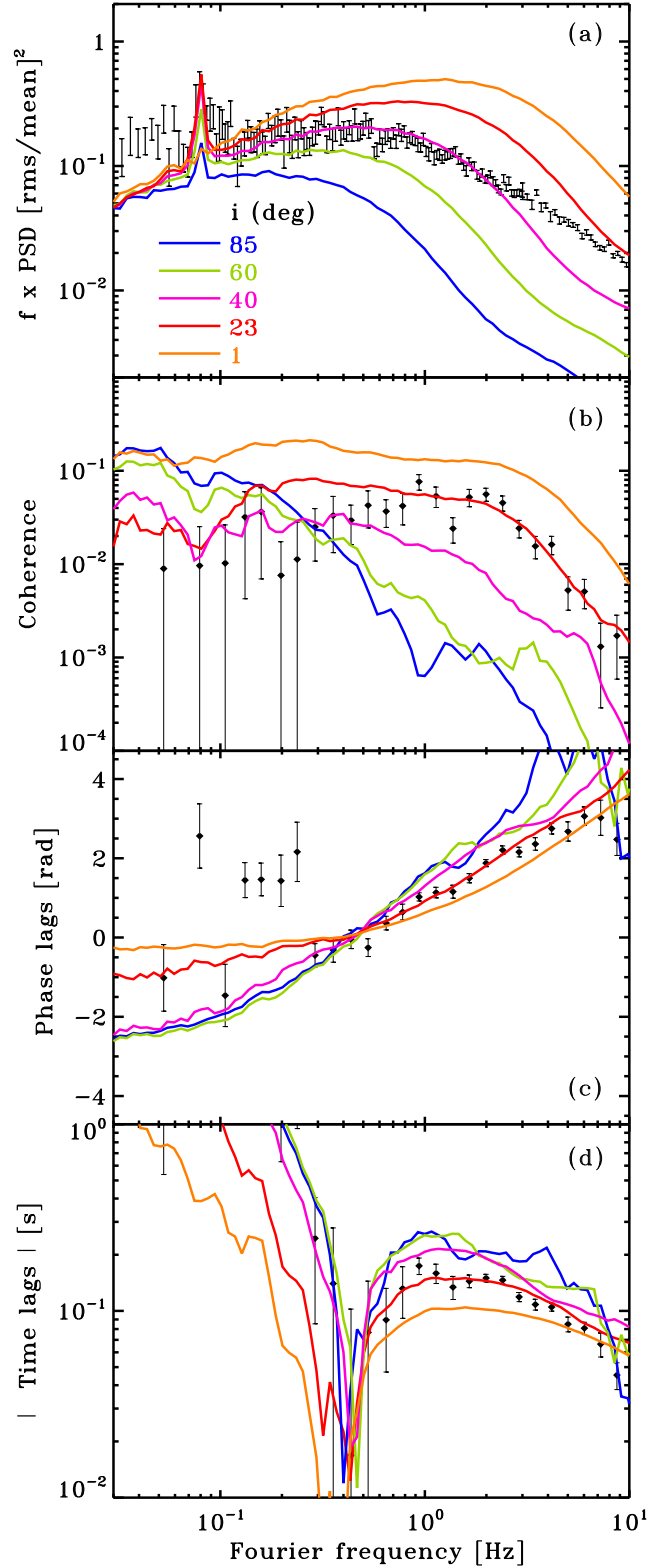}
   \caption{ Dependence of the timing characteristics on jet inclination  The model parameters are shown in Table~\ref{tab:par}, models C1 to C4 and B.}
     \label{fig:incl}
\end{figure}

\begin{figure}
	\includegraphics[width=\columnwidth]{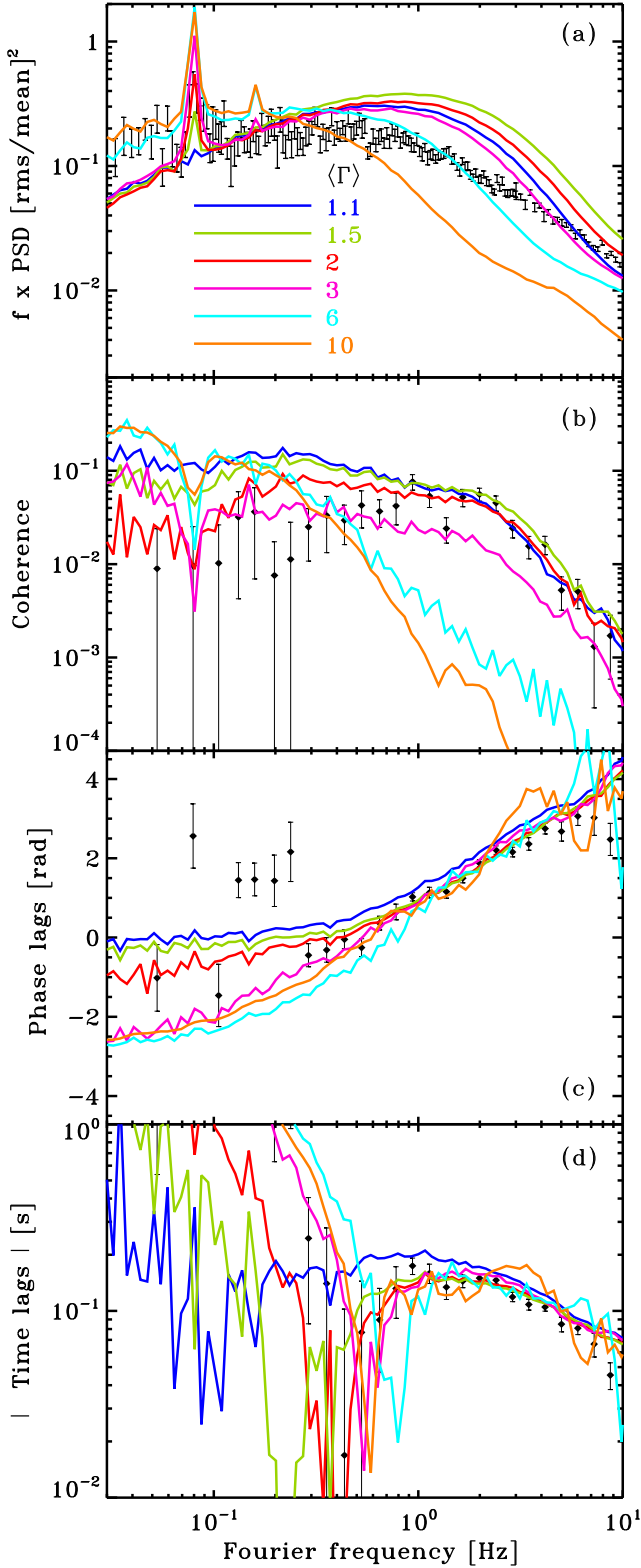}
   \caption{ Dependence of the timing characteristics on jet Lorentz factor. The  model parameters are shown in Table.~\ref{tab:par}, models D1 to D5  and B.}
     \label{fig:gam}
\end{figure}

 \begin{figure}
	\includegraphics[width=\columnwidth]{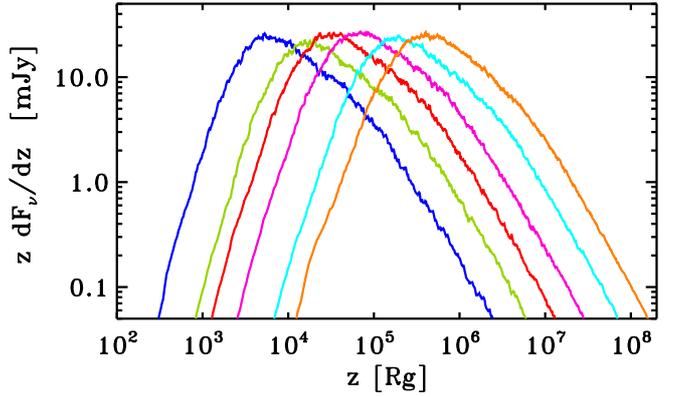}
   \caption{ IR emissivity profiles: dependence on jet Lorentz factor, the different models (colours) corresponds of those of Fig.~\ref{fig:gam}}
     \label{fig:gamprof}
\end{figure}

\begin{figure}
	
	\includegraphics[width=\columnwidth]{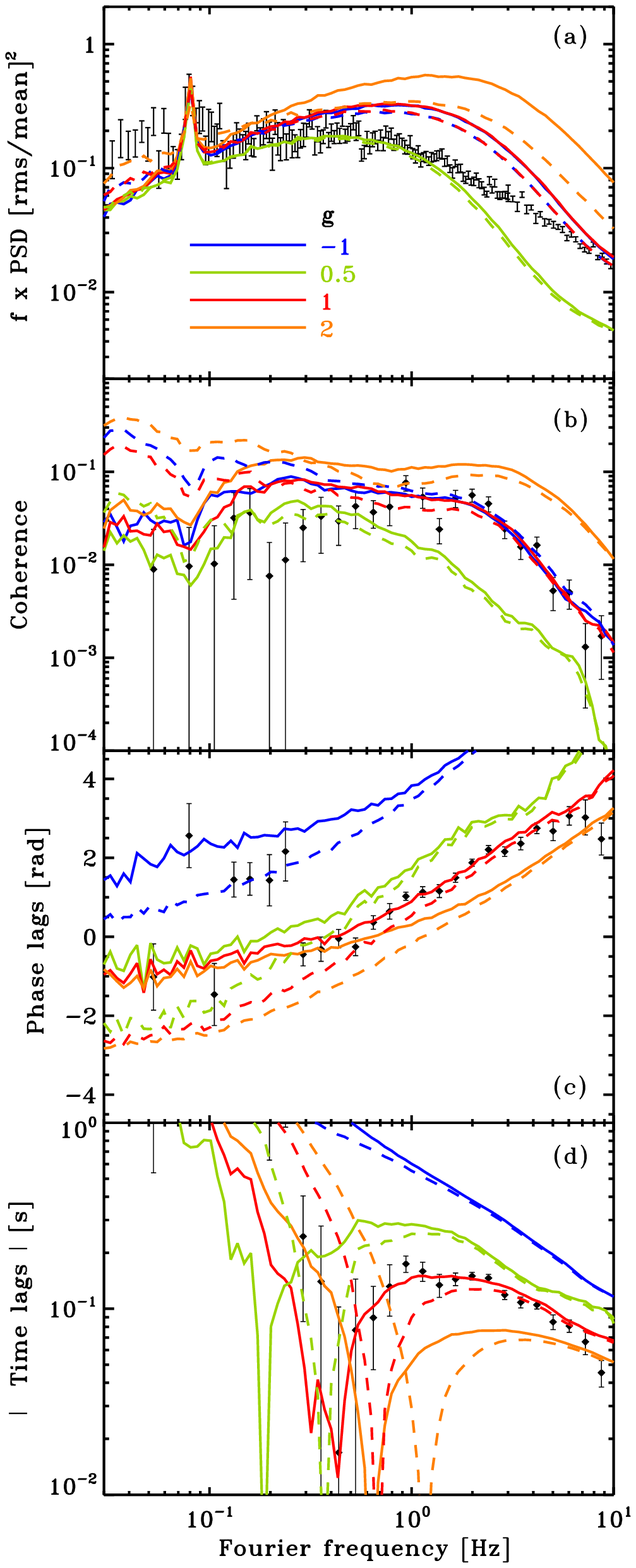}
   \caption{ Dependence of the timing characteristics on jet accretion coupling . The  model parameters are shown in Table.~\ref{tab:par}. The full lines show the results of simulation assuming that the mass of the ejected shells is a constant (models E1 to E3 and B). The dashed lines show the results of simulation assuming that the kinetic jet power does not vary in time (models F1 to F4).}
     \label{fig:gm0}
\end{figure}

The inclination angle $i$ does not affect the intrinsic dynamics of the shells collisions, but it changes the amplitude of the Doppler effects.  To illustrate the effects we start from fiducial model B  and varied  the  jet inclination $i$, in the range 1$^{\circ}$ to 85$^{\circ}$. For each value of the inclination, the jet kinetic power was also modified to maintain a good agreement with the spectral data.  The 'best fit' parameters are shown in Table~\ref{tab:par} (models C1 to C4). The timing results are shown in Fig.~\ref{fig:incl}. 

The apparent time-scales of the fluctuations increase with $i$ like $1-\bar{\beta}\cos i$, which explains the shift of the PSD break toward low Fourier frequencies at larger inclination that can be seen in the top panel of Fig.~\ref{fig:incl}.  The break frequency in the coherence spectrum follows the evolution of  the PSD break. 
Because the observed time-scales increase with $i$, the high frequency lags are longer at high inclination. However the shift of the lag spectrum towards lower frequencies combined with the slow decrease of the lag with frequency mean that the lag amplitude is not changed by more than a factor of 2 or so. Finally, the time-lags at high frequency are not very sensitive to $i$ and remains of the order of 100 ms.  

{ We  note the significant  changes in the amplitude of the low frequency negative phase lags which are close to 0 for face-on and approach  $-\pi$ for edge-on inclinations.  
This change is due to the different modulation of the IR light curves by the fluctuations of the relativistic Doppler boosting factor $\delta$ at different angles.  For $\sin{i}<1/\Gamma$, the relativistic Doppler boosting factor $\delta$ is correlated with Lorentz factor, $\Gamma$, of the ejecta.   Since we assumed the X-ray flux scales linearly with $\Gamma$,  this will add a positive correlation  between X-ray and IR  light curves which will reduce the low frequency anti-correlation  caused by the dips of $\Gamma(t)$ (see section~\ref{sec:coherenceandlags}), and enhance the positive response of both dips and spikes.  This results in the reduction of  the negative lags. 
On the contrary, for $\sin{i}>1/\Gamma$,  an anti-correlation is expected between $\Gamma$ and $\delta$ and this makes the lags more negative.}

Finally, as expected, the lower inclinations allow for a stronger modulation through jet precession which leads to a more prominent IR QPO. 

{  When qualitatively comparing the different models of Fig.~\ref{fig:incl} to the data, one can see that the the PSDs obtained for higher inclinations ($\ge 40^{\circ}$) are closer to the observed IR PSD shape than the prediction obtained for our fiducial inclination of  23$^{\circ}$. Indeed, they have a flatter low frequency component (in $f\times $PSD representation) and a break at lower frequency.  A higher inclination however also leads to a loss of coherence at high Fourier frequencies, this does not fit  the high-frequency coherence spectrum as well as the model with $i=23^{\circ}$. A higher inclination is also energetically demanding as it requires a larger jet power to fit the SED (see Table~\ref{tab:par}). For our fiducial average jet Lorentz factor $\bar{\Gamma}=2$, the trade-off between these effects suggests a preferred viewing angle  in the approximate range $20^{\circ}<i<40^{\circ}$.}

\subsection{Effects of jet Lorentz factor}\label{sec:ejlf}

Among the model parameters, the average jet Lorentz factor $\bar{\Gamma}$ is the only one which has direct effects on the dynamics of the collisions. 
 Again we start from fiducial  model B and check the effects of  the time averaged jet Lorentz factor. We calculated models for jet Lorentz factors in the range 1.1--10, for each value of the jet Lorentz factor the jet kinetic power and the jet opening angle are modified to maintain a good agreement with the spectral data.  The 'best fit' parameters are shown in Table~\ref{tab:par} (models D1 to D5). The timing results are shown in Fig.~\ref{fig:gam}. 

As $\bar{\Gamma}$ increases the apparent time-scales become faster due to the Doppler beaming effects as in the case of a decreased inclination. However at the same time, the shocks  gradually  take place on larger scales, at a larger distance from the black hole. Indeed, the amplitudes of fluctuations of $\Gamma-1$ have the same fractional amplitude as the X-ray flux. Therefore, when $\bar{\Gamma}$ is increased the amplitude of the fluctuations also increases linearly. However, as the velocity of the ejecta approaches the speed of light, the amplitude of the velocity fluctuations are limited and decrease. Smaller jet velocity fluctuations imply that the collisions occur at larger distances from the black hole and at larger scales.  Fig.~\ref{fig:gamprof}   shows the IR emissivity profiles along the jet for the different jet Lorentz factors.  The distance to the peak of the IR emission increases with $\bar{\Gamma}$, by more than 2 orders of magnitude. This counters the  Doppler effect and tends to shift the break of the PSD towards lower frequencies as $\bar{\Gamma}$ increases.

The dependence of the relativistic Doppler boosting factor on velocity and inclination implies that on long time-scales the IR flux is more correlated  with the jet Lorentz factor for jet inclinations with respect to the line of sight so that  $\cos i > \beta$.
or equivalently $\bar{\Gamma} < 1/\sin{i}$.   For the inclination of 23$^{\circ}$  of the fiducial model this corresponds to $\bar{\Gamma} < 2.56$.    Below this limit the jet emission is correlated with $\Gamma$ and we obtain a lag close to zero at low frequencies.  For larger $\bar{\Gamma}$ the IR emission is anti-correlated with the fluctuations of $\Gamma$ and a stronger anti-correlation is observed.   At low frequencies, the phase lag is getting gradually closer to -$\pi$.  
 
Finally we note that the modulation associated with jet precession becomes stronger at large $\bar{\Gamma}$ as expected from the estimates of section~\ref{sec:anes}, and its harmonic content increases as well. The second harmonic of the IR QPO becomes apparent for $\Gamma \ge 3$ in Fig~\ref{fig:gam}.

{ When comparing to the data, larger Lorentz factors predict IR PSDs that are closer to the observation (see e.g. model  D4 with $\bar{\Gamma}=6$). However the effect of larger Lorentz factors is also to reduce the coherence at high frequencies and the coherence spectra is clearly not reproduced for $\bar{\Gamma}\ge 3$.  If in addition, we consider that the harmonic of the IR QPO is not apparent in the data, we can conclude that our results indicate $\bar{\Gamma}\le 3$.
}

\subsection{Effects of jet disc coupling}\label{sec:jdc}

Of course the results depend also on the assumed relation between the instantaneous  X-ray flux and Lorentz factor.  So far we have assumed that $\Gamma-1$ scales linearly with the X-ray flux. We now  generalise this by assuming a non-linear relation:
\begin{equation}
\Gamma-1 \propto L_{\rm X}^g,
\end{equation}
where the fixed exponent $g$ can take any positive or negative value.

{ In practice we generate a time series $L_{\rm X}$ which has the same power spectrum as the observed X-ray light curve (as described in M14) and then we define the fluctuations of the Lorentz factor as:}
\begin{equation}
\Gamma=1+(\bar{\Gamma}-1)  \frac{L_{\rm X}^g}{\langle{L_{\rm X}^g}\rangle}
\end{equation}
{ where $\langle{L_{\rm X}^g}\rangle$ is the time average of $L_{\rm X}^g$. When $g$ differs from unity the rms amplitude of the fluctuations of $\Gamma-1$, and even their PSD shape can  strongly deviate from that observed in X-ray.} 

We calculated several models with different values of the $g$ parameter,  adjusting the jet power and opening angle in order to keep the SED in agreement with the data as indicated in Table~\ref{tab:par}. The results are shown in Fig.~\ref{fig:gm0}.

If  the exponent $g$ is positive, increasing $g$ increases the amplitude of jet Lorentz factor fluctuations. As a consequence the rms variability and the coherence increase, especially at high frequencies. Also the high frequency IR time-lag  decreases with increasing $g$  because the IR  emitting region is closer to the black hole (mostly due to the faster variability).
The choice of a negative  $g$ implies an anti-correlation between X-rays and jet Lorentz factor  $\Gamma$. The case  $g=-1$ is shown in Fig~\ref{fig:gm0}. The anti-correlation implies that the IR phase lag is shifted by an angle $\pi$ (compared to $g=1$).

 Fig.~\ref{fig:gm0} also displays similar models in which instead of assuming that the ejected shells have all the same mass, we assume that they all have the same kinetic power $(\Gamma-1)mc^2$, despite the fluctuations of $\Gamma$.  Assuming constant jet power amplifies the anti-correlation at low $f$ because an increase in jet Lorentz factor can lead to a larger decrease in IR flux due to the reduced mass/energy densities of the faster shells. On the other hand the high frequency lags are not significantly affected by the choice of this prescription.

Overall models with $g\simeq 1$ appear to provide better agreement with the data. 
We note that this may appear in contradiction with the results of  M14 who  presented synthetic IR light curves and showed that  {\sc ishem} could produce an IR vs X-ray cross-correlation function that is similar to the one observed by C10, including a similar $\sim100$~ms lags under the assumption of $g=-1$. However in this study the SED was not simultaneously fitted with {\sc ishem},  and the assumed injected fluctuations of $\Gamma$ were not fixed by the observed PSD but arbitrarily set with a flicker noise power spectrum  extending up to 50 Hz.  This causes the different results. In fact, the X-ray PSD of GX 339--4 in the hard state does not extend to such high frequencies. It  usually shows a break or cut-off below 10 Hz.

{  The linear $L_{\rm X}\propto  \Gamma-1$  relation favoured by the data  could be interpreted as follows.
Let us consider a radiatively efficient accretion flow in which a fraction $b$ of the available accretion power $P_{\rm ac}$ is used for the jet ($P_{\rm j}=bP_{\rm ac}$) while the remaining part is radiated with a luminosity $P_{\rm rad}$. Since the X-ray luminosity is a good tracer of $P_{\rm rad}$:
\begin{equation}
L_{\rm X} \sim P_{\rm rad} = (1-b) P_{\rm ac} =(1/b-1) P_J= (\Gamma-1) (1/b-1) \dot{M}_j c^2
\end{equation}
where  $\dot{M}_j $ is the mass ejection rate. 
In this case we see that a positive linear connection between X-ray luminosity and $\Gamma$-1 would happen if $(1/b-1) \dot{M}_j$ is a constant (or at least weakly variable on short time-scales).  In the context of  {\sc ishem}, this condition is realised in models in which the shells are ejected with a constant mass. Indeed, since we have a uniform ejection time-step,  $\dot{M_{J}}$ is contant when averaged over times-scales longer than the ejection time-step (which is shorter than the time-scales probed by our observations). Together with the linear ($g=1$) relation between X-ray luminosity and $\Gamma$, this also implies a constant $b$. }
\section {Conclusions}

 Overall our  results confirm that jet emission powered by internal shocks driven by the variability of the accretion flow are an excellent candidate for the radio to IR emissions of GX 339-4.  We have shown that the same model that reproduces the radio IR SED  of GX 339-4 predicts IR variability properties that are very similar to those observed in this source.  In particular the X-ray vs IR coherence and  Fourier lags spectrum are astonishingly well reproduced provided the X-ray flux scales linearly with the fluctuations of the jet Lorentz factor.  At high Fourier  frequency the variability of the IR light curves is driven  by shell collisions occurring close to the base of the jet emitting region. The shell travel time from the disc, as measured in the observers frame, corresponds to the  observed 100 ms IR lag.  At lower frequencies  the IR variability is dominated by the longer time-scale variations of the Lorentz factor. The long time-scale IR fluctuations are roughly related to the time-derivative of the Lorentz factor modulated by fluctuations of the Doppler boosting factor.   
  There is { not much} room for fine tuning of the model parameters as the observed X-ray PSD determines entirely the shape of the SED, and in large part the IR timing properties of the source. As long as the model parameters are set in order to fit the radio and IR fluxes, the timing properties remain mostly constant, unless the inclination of the jet or the average jet Lorentz factor are varied. The averaged jet Lorentz factor controls the distance scale at which shocks are produced in the jet and this affects the observed time-scales of the  variability. The time-scales of the observed IR fluctuations also depend on the inclination $i$ through changes in the Doppler factor.  Both inclination and jet Lorentz factor affect the low frequency modulation of the IR light curve through Doppler amplification effects.  Depending on $i$ and $\bar{\Gamma}$, the IR light curves and  $\Gamma(t)$ can be either correlated or anti-correlated. The latter could be the cause of the observed  negative IR vs X-ray phase lags  observed at low Fourier  frequencies in GX 339--4. On the other hand, the high frequency  IR lags are not dramatically  affected and remain of the order of 100 ms, unless the connection between X-ray flux and $\Gamma$ is strongly non-linear (i.e. $g$ differs from unity). In fact, the 100 ms time-scale is determined mostly by the high Fourier frequencies of the observed X-ray PSD used as input of our model.

We have shown that IR QPOs of amplitude comparable to that observed by K16 can be produced by jet precession provided that coherent precession is maintained during $\sim 10$ cycles.  { If the whole jet precessed with the hot flow, this could lead to much larger jet opening angles than observed. However, in the course of the multiple collisions encountered by the shells as they travel down the jet, their velocity vectors average to the direction of the precession axis. Jet precession can therefore  be maintained only close to the black hole and does not affect the large scale structure of the jet. As a corollary the model predicts that the amplitude of the QPO should decrease quickly with photon wavelength.}

We find the amplitude of the QPO and its harmonic content to be strongly dependent on the jet Lorentz factor and various other geometrical parameters. Future comparisons to data using a combination of accretion flow  and jet precession models for the  X-ray and IR QPOs could prove extremely constraining for the geometry of the accretion ejection system. The observables to be reproduced include not only the amplitude and profile of  both QPOs but also the IR vs X-ray phase lags at the QPO frequencies. In the case of the K16 data, the hot flow precession geometry must allow for a dominant X-ray QPO harmonic. It is far from given that all these features can be simultaneously reproduced in the framework of the precession model.  

Although there are many parameter degeneracies, our  modelling of the data from K16 suggests the jet is mildly relativistic at most. 
 This is indicated by the  absence of a detected harmonics of the IR QPOs. The comparisons of the model predictions to the observed Fourier coherence spectrum also suggest that the average  jet Lorentz factor $\bar{\Gamma}\le 3$ otherwise the predicted coherence is too low above 1 Hz. The jet average velocity may depend on the luminosity and spectral state as suggested by the recent study of P\'eault et al. (2018) who used {\sc ishem} to model the evolution of the SED of the black hole binary  MAXI~J1836-194 during an outburst and found that $\bar{\Gamma}$ had to be increased  from $\simeq1.1$ to $\simeq 10$  while the source evolved from the low-hard to the hard-intermediate X-ray state.  
  
Despite our successful  modelling  of the IR-X-ray correlations observed in GX 339-4 with  {\sc ishem}, we found that the predicted amplitude of IR variability is significantly larger than that observed.  This remains a problem for our model.  This may be solved by the presence of an additional constant component, originating either from the disc or the jet itself  and dominating the IR flux. It is also possible that radiative cooling, which is not taken into account in the present version of the model would damp the IR fluctuations to a level closer to that observed.  These effects will be studied in future works.

\section*{Acknowledgements}
This work has  been carried out also thanks to the support of the OCEVU Labex (ANR-11-LABX-0060) and the A*MIDEX project (ANR-11-IDEX-0001-02) funded by the "Investissements d'Avenir" French government program managed by the ANR.
It also received fundings from PNHE in France, and from the french Research National Agency : CHAOS project ANR-12-BS05-0009 (http://www.chaos-project.fr). 
 The authors thank Adam Ingram for useful discussions of the  LT precession model for X-ray QPOs and the anonymous referee for useful suggestions.

%%%%%%%%%%%%%%%%%%%%%%%%%%%%%%%%%%%%%%%%%%%%%%%%%%



%\bibliographystyle{mnras}


%%%%%%%%%%%%%%%%%%%%%%%%%%%%%%%%%%%%%%%%%%%%%%%%%%

%%%%%%%%%%%%%%%%% APPENDICES %%%%%%%%%%%%%%%%%%%%%

\appendix

\section{Spectral fits combining {\sc ishem} and {\sc diskir}}\label{sec:diskir}

 \begin{table*}
 \caption{Diskir model parameters}
 \label{tab:pardir}
 \begin{tabular}{cccccccccc}
  \hline 
  {\sc ishem } Model & $N_{\rm H}$ (10$^{21}$ cm$^{-2}$) &$kT_{\rm disc}$ (keV) & $\Gamma$ & $kT_{\rm e}$ (keV)& $L_{\rm c}/L_{\rm d}$ & $f_{\rm out}$ & log$R_{\rm out}$ &  $N_{\rm diskir}$   & $\chi^2$/d.o.f  \\ 
  \hline 
 A & 6.1$\pm$0.2 &0.142$^{+0.003}_{-0.005}$  &    1.78 $\pm$0.02    &     14.1$^{+3.2}_{-1.7}$          &     2.61$^{+0.82}_{-0.46}$        &  5$\times 10^{-2}$ (f)   &  3.19$^{+0.27}_{-1.19}$ & 7.10$^{+0.81}_{-0.77}$ $\times10^5$  & 1817/1952    \\
 B   &   6.1  $\pm 0.2$   &        0.142$^{+0.003}_{-0.005}$        &     1.78 $\pm$0.02                &        14.1$^{+3.2}_{-1.7}$               &       2.60$^{+0.80}_{-0.46}$   & 5$\times 10^{-2}$ (f) & 3.01$^{+0.37}_{-1.25}$& 7.18$^{+0.84}_{-0.76}$ $\times10^{5}$  & 1830/1952 \\    
 A' & 5.2$\pm$0.1 &0.116$^{+0.003}_{-0.002}$  &    1.75 $\pm$0.02    &     21.2$^{+3.2}_{-1.7}$          &     9.86$^{+0.82}_{-0.46}$        &  5$\times 10^{-3}$ (f)   &  4.81$^{+0.27}_{-1.19}$ & 4.80$^{+0.81}_{-0.77}$ $\times10^5$  & 1797/1952    \\
                 
  \hline
 \end{tabular}
\end{table*}

We fit the observed SED of GX 339-4 using {\sc xspec} (Arnaud 1996), with a model combining {\sc ishem}, a self-irradiated accretion flow model {\sc diskir} (Gierli{\'n}ski et al. 2009), a reflection component {\sc pexrav} (Magdziarz \& Zdziarski 1995),  a gaussian line to model  the Fe K line around 6.4 keV, and neutral X-ray absorption ({\sc tbabs} model in {\sc xspec}).

In order to fit with the {\sc ishem} model in {\sc xspec} we have written a simple `local model' routine that uses a pre-calculated synthetic jet SED resulting from a specific {\sc ishem} simulation. The {\sc xspec} routine attempts to match the data by shifting the pre-calculated model in frequency and normalization. As the emission from the higher energy end of the particle distribution is not treated in the current version of {\sc ishem}, the {\sc xspec} model offers the possibility to add { an exponential cut-off} at  high frequency to the optically thin synchrotron emission from the jet.  The three free  parameters of the model are the shift in frequency, the shift in normalization and the cut-off energy $E_{\rm cut}$. Once the shift in frequency and normalization required to fit the data is determined from the fit, we can use it to calculate analytically the different combinations of {\sc ishem} parameters that would allow us to produce such a fit (see P\'eault et al. 2018 for details). This is how the model parameters listed in Table~\ref{tab:par} were determined. Once a parameter set is chosen,  we can  run a new simulation and fit again with {\sc xspec} to check that this best fit parameter set requires negligible shift in frequency and normalization to match the data.

The {\sc diskir} model calculates the emission of a truncated accretion disc irradiated by a hot Comptonizing accretion flow.  We found that reasonable fits of the XMM-Newton and RXTE data of K16 also require an iron line and reflection component. However the many parameters of the {\sc diskir}  and reflection are degenerate and we decided to 'freeze' many of them in our fits.  
We fixed the gaussian line energy and width at 6.7 keV and 0.5 keV respectively. In {\sc pexrav}  the primary emission parameters were  tied to the {\sc diskir} parameters of the Comptonizing plamas (spectral index $\Gamma$ and temperature $kT_e$), the inclination was fixed to the jet inclination of {\sc ishem} (i.e. 23$^{\circ}$ in most cases), the abundances are assumed to be solar and the reflection coefficient $R$ was set to -1 so that the primary emission is ignored and {\sc pexrav}  returns a pure reflection component. To account for the reflection and iron line we have only  two remaining free parameters which are their respective normalizations. In addition the {\sc diskir} parameters related to the irradiation of the inner disc where also fixed at their default values in our fits, namely the fraction of Comptonized radiation reprocessed in the inner disc was set to $f_{\rm in}=0.1$ , and the radius of the Compton illuminated disc in terms of the inner disc radius was set to  $R_{\rm ir}=1.2$.   Due to absolute calibration errors, fitting simultaneously the RXTE and XMM-Newton requires a different normalization. In order to account for this uncertainty, we  choose to multiply the whole model by a normalization constant when comparing it to the XMM-Newton spectrum. This constant is a fitted parameter that we find to be close to 0.75 in all of our fits. Note that the XMM-Newton data shown in Figs.~\ref{fig:fidsed} and~\ref{fig:fidsedbigshift} were corrected by this factor to match the RXTE data.

We also found that  due to the relatively  poor quality of our IR to UV coverage it was not possible to constrain simultaneously the shift and normalization of  {\sc ishem} and the reprocessed emission originating in the outer accretion disc.  The latter is mostly controlled by the parameter $f_{\rm out}$, the fraction of Comptonized radiation that is reprocessed in the outer disc, and  $R_{\rm out}$ the outer disc radius in terms of the inner disc radius.  We therefore set a fixed value of $f_{\rm out}$ in our fits.
And, since we are mostly interested in models in which the jet dominates the IR emission, we first fitted the radio to IR SED with {\sc ishem} only. Then, keeping the {\sc ishem} parameters fixed, we added the optical/UV and X-ray data and fitted for the accretion flow parameters.  The best  fits that we obtained for {\sc ishem} simulations using the observed X-ray PSD of GX 339-4 as input, including the QPO (model A), or not (model B)  were statistically acceptable with reduced $\chi^2<1$. The best fit model parameters of the {\sc diskir} model corresponding to {\sc ishem} models A and B are very similar and are shown in Table~\ref{tab:pardir}.

We note that due to the model degeneracy as well as the gaps in our broadband coverage, its possible to find different model parameters that provide a statistically  equivalent  representation of the data. In particular, as a possible solution to the strong IR variability predicted by {\sc ishem} we can find a spectral decomposition of the observed SED in which the IR emission is dominated by the accretion flow. Changing the jet power and jet opening angle of model A to $P_J=0.30$ and $\phi=20^{\circ}$, the jet SED shifts redwards in frequency by a factor 0.24 and the model normalization is reduced by a factor 0.55.  Alternatively, increasing $\bar{\Gamma}$ from 2 to 3, we would obtain an identical SED for $P_J=0.21$ and $\phi=4.8^{\circ}$. The IR flux predicted by this model (hereafter model A') represents less than 20 percent of the observed one. Then freezing  $f_{\rm out}=5\times10^{-3}$, a {\sc diskir} model fit allows to account for the IR to X-ray emission. The result is displayed on Fig.~\ref{fig:fidsedbigshift} and the best fit parameters are shown in  Table~\ref{tab:pardir}.

%%%%%%%%%%%%%%%%%%%%%%%%%%%%%%%%%%%%%%%%%%%%%%%%%%

% Don't change these lines
\bsp	% typesetting comment
\label{lastpage}
\end{document}